\documentclass[aps, prx, 10pt, twocolumn,floatfix, nofootinbib]{revtex4-2}
\usepackage{braket}
\usepackage{soul}
\usepackage{amsfonts, amssymb, amsmath, mathtools, bm}
\usepackage{siunitx}
\usepackage{booktabs}
\usepackage[dvipsnames]{xcolor}
\usepackage{xurl}
\usepackage[breaklinks=true,colorlinks,citecolor=blue,linkcolor=blue,urlcolor=blue]{hyperref}
\usepackage{comment}
\usepackage{makecell}
\usepackage{multirow}
\usepackage{orcidlink}
\usepackage[capitalize]{cleveref}

\newcommand{\trace}[1]{\text{tr}#1}

\begin{document}

\title{Protein folding with an all-to-all trapped-ion quantum computer}

\author{Sebastián V. Romero$^{\orcidlink{0000-0002-4675-4452}1,2}$}
\author{Alejandro Gomez Cadavid$^{\orcidlink{0000-0003-3271-4684}1,2}$}
\author{Pavle Nika\v{c}evi\'{c}$^{\orcidlink{0000-0002-8832-5541}1}$}  

\author{Enrique Solano$^{\orcidlink{0000-0002-8602-1181}1}$}
\email[]{enr.solano@gmail.com}
\author{Narendra N. Hegade$^{\orcidlink{0000-0002-9673-2833}1}$}
\email[]{narendrahegade5@gmail.com}
\affiliation{$^{1}$Kipu Quantum GmbH, Greifswalderstrasse 212, 10405 Berlin, Germany\\$^{2}$Department of Physical Chemistry, University of the Basque Country UPV/EHU, Apartado 644, 48080 Bilbao, Spain}

\author{Miguel Angel Lopez-Ruiz$^{\orcidlink{0000-0002-8152-5655}}$}
\affiliation{IonQ Inc., 4505 Campus Dr, College Park, MD 20740, USA}

\author{Claudio Girotto$^{\orcidlink{0000-0001-8739-5866}}$}
\affiliation{IonQ Inc., 4505 Campus Dr, College Park, MD 20740, USA}

\author{Masako Yamada$^{\orcidlink{0009-0004-0388-3053}}$}
\affiliation{IonQ Inc., 4505 Campus Dr, College Park, MD 20740, USA}

\author{Panagiotis Kl. Barkoutsos$^{\orcidlink{0000-0001-9428-913X}}$}
\affiliation{IonQ Inc., 4505 Campus Dr, College Park, MD 20740, USA}

\author{Ananth Kaushik$^{\orcidlink{0009-0009-2799-1194}}$}
\email{kaushik@ionq.co}
\affiliation{IonQ Inc., 4505 Campus Dr, College Park, MD 20740, USA}

\author{Martin Roetteler$^{\orcidlink{0000-0003-0234-2496}}$}
\email{martin.roetteler@ionq.co}
\affiliation{IonQ Inc., 4505 Campus Dr, College Park, MD 20740, USA}

\date{\today}

\begin{abstract}
\noindent
    We experimentally demonstrate that the bias-field digitized counterdiabatic quantum optimization (BF-DCQO) algorithm, implemented on IonQ’s fully connected trapped-ion quantum processors, offers an efficient approach to solving dense higher-order unconstrained binary optimization (HUBO) problems. Specifically, we tackle protein folding on a tetrahedral lattice for up to 12 amino acids, representing the largest quantum hardware implementations of protein folding problems reported to date. Additionally, we address MAX 4-SAT instances at the computational phase transition and fully connected spin-glass problems using all 36 available qubits. Across all considered cases, our method consistently achieves optimal solutions, highlighting the powerful synergy between non-variational quantum optimization approaches and the intrinsic all-to-all connectivity of trapped-ion architectures. Given the expected scalability of trapped-ion quantum systems, BF-DCQO represents a promising pathway toward practical quantum advantage for dense HUBO problems with significant industrial and scientific relevance.
\end{abstract}

\maketitle

\section{Introduction}
\noindent
Protein folding, a fundamental biological process, poses a grand challenge in scientific and industrial applications due to its immense medical relevance, particularly in drug discovery and disease understanding. From a computational perspective, protein folding is often framed as a search for the lowest-energy configuration on a lattice, which inherently maps to finding the ground state of a general $p$-spin-glass model. This formulation places it squarely within the class of Higher-Order Unconstrained Binary Optimization (HUBO) problems.

Combinatorial optimization problems, including protein folding, appear frequently across numerous scientific and industrial domains, ranging from logistics and transportation to chemistry and biology. These problems typically require finding optimal or near-optimal solutions within vast configuration spaces. Many, like HUBOs encountered in protein folding, are NP-hard, meaning that, in the worst case, the computational time required to solve them scales exponentially with the system size. While various classical solvers exist, including problem-specific or general-purpose algorithms such as simulated annealing~\cite{kirkpatrick1983optimization}, tabu search~\cite{tabu}, CPLEX~\cite{cplex}, and Gurobi~\cite{gurobi}, they often struggle with the non-convex energy landscapes characteristic of these complex problems, particularly in higher-order binary optimization~\cite{bertsekas2003convex, nonlinearprog, surveynonconvex, arjevani2020nonconvexchallenges, danilova2022nonconvexchallenges, xu2024relaxationschallenges}.

These optimization problems can be reformulated as energy minimization tasks over spin-glass models, whose ground states encode the optimal solutions~\cite{lucas2014ising}. Quadratic unconstrained binary optimization (QUBO) problems naturally map to Ising models, while their higher-order (HUBO) counterparts can be mapped into general $p$-spin-glasses~\cite{gardner1985spin}, also known as higher-order Ising models in quantum optimization~\cite{pelofske2023quantum, pelofske2024short-depth}. As problem complexity increases and classical resources become limiting, quantum computing emerges as a promising alternative. Quantum algorithms such as adiabatic quantum optimization (AQO)~\cite{albash2018adiabatic} and the quantum approximate optimization algorithm (QAOA)~\cite{farhi2014quantumapproximateoptimizationalgorithm}, a digitized form of adiabatic quantum computing~\cite{barends2016digitized}, offer a new approach to tackle computationally intractable optimization problems~\cite{abbas2024challenges, boulebnane2022solving, kotil2025quantum, koch2025quantum}.

Breakthrough developments in quantum technologies are making quantum computers increasingly capable of addressing real-world optimization problems. In parallel, quantum algorithms have been proposed with the potential for theoretical speedups over classical solvers~\cite{durr1999quantumalgorithmfindingminimum, somma2008quantum, wocjan2008speedup, hastings2018shortpathquantum, montanaro2018quantum-walk, montanaro2020quantumspeedup, chakrabarti2022universalquantumspeedupbranchandbound, dalzell2023mind}. QAOA, in particular, has garnered attention as a versatile technique for large-scale optimization problems, showing compelling evidence of quantum advantage in combinatorial optimization. For example, benchmark experiments on 8-SAT instances up to 20 variables~\cite{boulebnane2022solving} and the low autocorrelation binary sequence (LABS) problem~\cite{boehmer1967binary, schroeder1970synthesis, shaydulin2024evidence} have demonstrated that QAOA can outperform classical solvers in terms of time to solution, highlighting the practical relevance of quantum algorithms in optimization. However, HUBO problems are considerably harder to solve than QUBO problems and frequently appear in applications like the LABS problem~\cite{shaydulin2024evidence, koch2025quantum}, factorization~\cite{hegade2021factorization}, and satisfiability problems~\cite{Battiti2009}. Despite recent experimental implementations of HUBO problems on quantum hardware~\cite{pelofske2023quantum, pelofske2024short-depth, barron2023provableboundsnoisefreeexpectation}, their implementation is particularly challenging due to current hardware limitations, such as qubit connectivity, decoherence, and gate fidelities. 

Recent advances, specifically targeting the protein folding problem, include quantum annealing~\cite{perdomo2008construction,perdomo2012finding,babbush2014construction,babej2018coarsegrainedlatticeproteinfolding} and variational quantum algorithms~\cite{robert2021resource,chandarana2023digitized,pamidimukkala2024protein,wang2025efficient}. The limitations of these models include the approximate treatment of chemical interactions and little flexibility in the possible geometric structure since the amino acid positions are constrained to the imposed lattice. Nevertheless, as long as the obtained optimal local structure of a studied protein domain (or an oligopeptide) is within the ``funnel'' leading to the global minimum, the accurate folded protein structure can be easily reached using the standard atomistic geometry optimization techniques starting from the obtained structure as the initial geometry. Additionally, the high density of the associated $p$-spin-glass makes it challenging to implement time-evolution-based quantum algorithms on limited connectivity architectures due to swap overheads. 

Counterdiabatic (CD) protocols~\cite{demirplak2003adiabatic, berry2009transitionless, chen2010fast, campo2013shortcuts,  sels2017minimizing, claeys2019floquet, takahashi2024shortcuts}, designed to suppress undesired diabatic transitions, have shown superior performance in optimization tasks. In addition to digitized counterdiababtic QAOA~\cite{chandarana2022digitized} and digitized counterdiababtic quantum optimization algorithm (DCQO)~\cite{hegade2022digitized}, as well as the recent branch-and-bound DCQO~\cite{simen2025branch}, the bias-field DCQO (BF-DCQO) algorithm~\cite{cadavid2024bias, romero2024bias, iskay} combines concepts from quantum annealing with dynamically updated bias fields~\cite{grass2019quantum, grass2022quantum}. BF-DCQO evolves the system towards optimal solutions without relying on variational methods that can suffer from barren plateaus~\cite{cerezo2024doesprovableabsencebarren,larocca2024reviewbarrenplateausvariational}, resulting in a purely quantum and scalable algorithm for HUBO problems~\cite{romero2024bias}. Recently, it has been demonstrated that BF-DCQO exhibits a runtime quantum advantage compared to classical solvers in achieving approximate solutions for specific HUBO problems \cite{chandarana2025runtime}.

In this work, we combine the BF-DCQO algorithm’s resource efficiency and fast convergence with IonQ’s trapped-ion hardware, which offers all-to-all connectivity and high gate fidelities~\cite{ionq}. Using this setup, we tackle the protein folding problem for sequences of up to 12 amino acids mapped onto 33 qubits. To further demonstrate the versatility of this approach, we solve fully connected Ising spin-glass problems~\cite{sherrington1975solvable} and MAX $k$-SAT instances~\cite{Battiti2009} on systems with up to 36 qubits.

The rest of the article is structured as follows: \cref{sec:formulation} presents a concise review of methods for tackling HUBO problems, introduces the BF-DCQO algorithm, and details the parameters used in the experiments. \cref{sec:problems} introduces the problems to be solved and how the instances were generated. \cref{sec:results} provides a detailed discussion of the IonQ Forte and IonQ Forte Enterprise quantum computers, and presents the BF-DCQO results obtained on the quantum hardware, comparing them with classical results. Finally, \cref{sec:conclusion} summarizes the conclusions drawn from the presented results. Further details and extended results are available in the Appendices.

\section{Higher-order quantum optimization: review of methods}\label{sec:formulation}

\subsection{Formulation}
\noindent
Numerous optimization problems relevant to both academia and industry can be represented as HUBO problems. By mapping each binary variable $x_i$ into a spin variable $s_i$ using the transformation $x_i = (1-s_i)/2$, a minimization problem is converted into finding the ground state of a $p$-spin-glass~\cite{lucas2014ising}, as described by the Hamiltonian
\begin{equation}\label{eq:hubo}
    H_f = \sum_{k=1}^p \sum_{\bm{i}\in G_{k}}J_{\bm{i}}\sigma^z_{i_1}\cdots \sigma^z_{i_k},
\end{equation}
where $\bm{i}\coloneqq (i_1,\cdots, i_k)$ and $G_k$ is a hypergraph containing the indices of the $k$-body terms, each weighted by the coupling strengths $J_{\bm{i}}$, with $k\in[1,p]$.

Although a HUBO problem can be converted into a QUBO formulation, it introduces additional variables and constraints that are not inherent to the original problem. This can increase the complexity and resource demands of the optimization task. Additionally, classical solvers often exploit the convexity of their quadratic counterparts, which is not usually guaranteed since common convex relaxation techniques prove ineffective~\cite{bertsekas2003convex, nonlinearprog, surveynonconvex, arjevani2020nonconvexchallenges, danilova2022nonconvexchallenges, xu2024relaxationschallenges}. In contrast, quantum hardware is increasingly capable of natively supporting higher-order interactions, potentially overcoming these challenges.

One prominent quantum optimization method, AQO, relies on the adiabatic theorem. It aims to adiabatically evolve a system from a readily preparable initial ground state to the ground state of the target problem Hamiltonian $H_f$,  \cref{eq:hubo}. This evolution can be described as $H_{\text{ad}}(\lambda)=  (1-\lambda) H_i + \lambda H_f$, where $\lambda(t)$ is a scheduling function that toggles from zero to one in the time interval $t\in[0,T]$. In the adiabatic limit $\dot{\lambda}(t)\to 0$, the system reaches the ground state of $H_f$. The initial Hamiltonian is typically $H_i=\sum_j(h^x_j\sigma^x_j + h^b_j\sigma^z_j)$, whose ground state is known for any value of the transverse, $h^x_j$, and longitudinal, $h^b_j$, field contributions acting on the $j$th spin. However, implementing higher-order terms in analog platforms remains challenging, as HUBO-to-QUBO maps are still needed, as seen in quantum annealers~\cite{dwave}.
On the other hand, digital quantum platforms offer more flexibility due to their universality, allowing higher-order terms to be decomposed into products of one- and two-qubit gates. Most of the digital approaches make use of variational quantum algorithms to address HUBO problems~\cite{robert2021resource,boulebnane2022solving,chandarana2023digitized,pelofske2023highround,pelofske2023quantum,pelofske2024short-depth,shaydulin2024evidence,wang2025efficient}. However, these methods are known to suffer from barren plateaus, and their scalability is still under debate~\cite{cerezo2024doesprovableabsencebarren,larocca2024reviewbarrenplateausvariational}.

\subsection{Digitized counterdiabatic quantum optimization}
\noindent
By introducing a counterdiabatic term, is possible to suppress diabatic transitions that happen when evolving a system for a finite time following an adiabatic Hamiltonian $H_{\rm ad}$~\cite{demirplak2003adiabatic, berry2009transitionless}. This term takes the form 
\begin{equation}
    H_\text{cd}(\lambda)=H_\text{ad}(\lambda)+\dot{\lambda}A_\lambda,
\end{equation}
with $A_\lambda$ the adiabatic gauge potential (AGP)~\cite{kolodrubetz2017geometry} and $\lambda(t)$ is a schedule for the evolution of the system. Nevertheless, its implementation is impractical due to its many-body structure and the need to know the entire spectrum. Accordingly, approximate implementations have been proposed~\cite{kolodrubetz2017geometry,sels2017minimizing,claeys2019floquet,hatomura2021controlling,takahashi2024shortcuts}, making it possible to expand the gauge potential in a nested-commutator series up to order $l$ as
\begin{equation}
    A^{(l)}_\lambda=i\sum_{k=1}^l \alpha_k(\lambda) \mathcal{O}_{2k-1} (\lambda)   
\end{equation}
with $\mathcal{O}_{0}(\lambda) = \partial_\lambda H_{\text{ad}}$ and $\mathcal{O}_{k}(\lambda) = [ H_{\text{ad}}, \mathcal{O}_{k-1}(\lambda) ]$. In the limit $l\to\infty$, it converges to the exact gauge potential. The coefficients $\alpha_k$ are obtained by minimizing the action $S_l=\trace{[G_l^2]}$ with $G_l=\partial_\lambda H_\text{ad} - i\big[H_\text{ad},A^{(l)}_\lambda\big]$. For simplicity, we set $l=1$ in our studies and use natural units ($\hbar\equiv 1$). Using \cref{eq:hubo}, its first-order nested-commutator expansion reads as
\begin{equation}
    A^{(1)}_\lambda = -2i\alpha_1(t) \sum_{k=1}^p \sum_{\bm{i}\in G_{k}} \sum_{i_j\in\bm{i}} h^x_{i_j}J_{\bm{i}}\sigma^z_{i_1}\cdots\sigma^y_{i_j}\cdots \sigma^z_{i_k}.
\end{equation}

In the fast-evolution regime, the unitary operator corresponding to the adiabatic term, $H_\text{ad}$, can be effectively omitted, which reduces the resources required for time evolution. Evolving such systems in time on analog quantum platforms remains a challenge, primally due to the inherent non-stoquasticity~\cite{hormozi2017nonstoquastic} of the counterdiabatic terms. To overcome this issue, digitized counterdiabatic quantum protocols have been proposed for digital quantum computers~\cite{hegade2021shortcuts}. The resulting time-evolution operator can be decomposed, up to $n_\text{trot}$ Trotter steps, as
\begin{equation}
    U(T) \approx \prod_{k=1}^{n_{\text{trot}}} \prod_{j=1}^{n_\text{terms}} \exp [-i \Delta t \gamma_j(k \Delta t) H_j],    
\end{equation}
with $H_\text{cd}=\sum^{n_\text{terms}}_{j=1}\gamma_j(t)H_j$ decomposed into $n_\text{terms}$ different $H_j$ operators and $\Delta t=T/n_\text{trot}$. Here, we use $T=1$, and a single Trotter step is implemented to preserve fidelity.
Additionally, we set a gate-angle cutoff, $\theta_\text{cutoff}$, such that terms where $|\gamma_j(k\Delta t)\Delta t|\text{ mod }2\pi<\theta_\text{cutoff}$ are pruned due to its presumed negligible contribution, which lowers the amount of resources required but does reduce circuit expressivity. In particular, throughout our study, we consider two different pruning strategies, namely ``soft" and ``hard", with which to reduce the number of resources. The ``hard'' pruning strategy removes a larger number of terms than the ``soft'' pruning strategy. The values of $\theta_\text{cutoff}$ used are reported in~\cref{sec:problems}, which are set such that the corresponding circuits require on the order of several hundreds of entangling gates for both pruning strategies (see also \cref{app:pruning}).

Building upon the DCQO protocol~\cite{hegade2022digitized} and quantum annealing with bias fields~\cite{grass2019quantum,grass2022quantum}, BF-DCQO iteratively performs DCQO, taking the solution from each step as an input of the subsequent iteration~\cite{cadavid2024bias,romero2024bias,iskay}. In particular, we initially set $h^x_j=-1$ and $h^b_j=0$ such that $H_i=-\sum^{N-1}_{j=0}\sigma^x_j$, whose ground state becomes $\ket{\psi(0)}=\ket{+}^{\otimes N}$ with $\sigma^x\ket{\pm}=\pm\ket{\pm}$. We use a schedule for the evolution $\lambda(t)=\sin^2(\frac{\pi}{2}\sin^2\frac{\pi t}{2T})$. We update the initial Hamiltonian after each iteration as 
\begin{equation}\label{eq:updated_hi}
    \tilde{H}_i=H_i + \sum_{j=0}^{N-1}h^b_j(\braket{\sigma^z_j})\sigma^z_j,
\end{equation}
with $h^b_j(\braket{\sigma^z_j})$ a function applied to the expectation value $\braket{\sigma^z_j}$ obtained by sampling the lowest energy-valued solutions previously obtained, using a fraction $\alpha\in(0,1]$ of the lowest-energy samples $E_k$ from $E(\alpha)= (1/\lceil \alpha n_\text{shots}\rceil)\sum_{k=1}^{\lceil \alpha n_\text{shots}\rceil} E_k$, with $E_k\le E_{k+1}$~\cite{Barkoutsos2020improving,barron2023provableboundsnoisefreeexpectation,romero2024bias}. 

After the bias fields are updated, the ground state of the Hamiltonian in  \cref{eq:updated_hi} changes and must be prepared for the next circuit execution. The lowest eigenvalue of each single-body operator, $\left(h_i^x \sigma^x_i - h_i^b \sigma^z_i \right )$, is given by $\lambda^{\min}_i = -\sqrt{(h^b_i)^2 + (h^x_i)^2}$, with the corresponding eigenstate being $\ket{\tilde{\phi}}_i = R_y(\theta_i) \ket{0}_i$, where $\theta_i = 2\tan^{-1}\left[(h^b_i + \lambda^{\min}_i)/{h^x_i}\right]$ and $R_y(\theta)=\exp(-i\theta\sigma^y/2)$. Therefore, the ground state of $\tilde{H}_i$ can be prepared by applying $N$ single-qubit rotations around the $y$-axis as
\begin{equation}
    \ket{\tilde{\psi}_i} = \bigotimes_{i=1}^{N} \ket{\tilde{\phi}}_i = \bigotimes_{i=1}^{N} R_y(\theta_i)\ket{0}_i.
\end{equation}

\section{Problems addressed}\label{sec:problems}

\subsection{Protein folding}
\noindent
Proteins, essential for biological functions such as the catalysis of metabolic reactions and DNA replication, are complex chains of amino acids. Determining how a sequence of amino acids folds into a three-dimensional structure is a central problem in biochemistry, particularly as misfolding is linked to neurological diseases like Alzheimer’s and Parkinson’s. Levinthal's paradox highlights the complexity of this process, noting that random folding would take longer than the age of the universe, yet proteins fold rapidly~\cite{levinthal1968pathways,levinthal1969fold}. This implies that protein folding follows specific, guided dynamics that are not yet fully understood at scale. However, due to the immense computational cost of simulating the underlying biophysical conditions leading to the correct protein folding, researchers often employ heuristic methods, such as machine learning techniques, or map the folding process onto more tractable optimization problems. In the machine learning approach, AlphaFold has emerged as an indispensable tool to elucidate the three-dimensional structures of proteins found in nature \cite{jumper2021highly, evans2024alphafold3}. However, since AlphaFold is trained on evolutionary patterns in naturally found proteins, it is unreliable in various types of synthetic proteins and oligopeptides, which are highly important in modern pharmaceutical development \cite{wang2022therapeutic, dunkelmann2024adding, miura2023vitro}. 

Modeling protein folding as an optimization problem usually involves a search for the lowest-energy conformation on a two- or three-dimensional lattice.  This can be reformulated as a quantum ground-state problem, making it amenable to quantum computation. Recent proposals include using quantum annealing~\cite{perdomo2008construction,perdomo2012finding,babbush2014construction,babej2018coarsegrainedlatticeproteinfolding} and variational methods~\cite{robert2021resource,chandarana2023digitized,pamidimukkala2024protein,wang2025efficient}. In this work, we employ the model presented in Ref.~\cite{robert2021resource}, where amino acids are arranged in a tetrahedral lattice. The protein chain is constructed sequentially; starting from an arbitrary lattice point that represents the first amino acid in the sequence, each subsequent amino acid is located at one of the four neighboring lattice sites. We use a dense encoding scheme, where the relative position of each amino acid with respect to the previous one (``turn'') is encoded using two qubits. Interactions between amino acids are computed as Miyazawa-Jernigan contact energies~\cite{miyazawa1996residue}, considering only contacts between first-nearest neighbors. Its corresponding Hamiltonian reads as
\begin{equation}\label{eq:protein}
    \begin{split}
        H_f &= \text{const.} + \sum_i h_i \sigma^z_i + \sum_{i<j} J_{ij} \sigma^z_i \sigma^z_j \\
        &+ \sum_{i<j<k} K_{ijk} \sigma^z_i \sigma^z_j \sigma^z_k + \sum_{i<j<k<l} L_{ijkl} \sigma^z_i \sigma^z_j \sigma^z_k \sigma^z_l \\
        &+ \sum_{i<j<k<l<m} M_{ijklm} \sigma^z_i \sigma^z_j \sigma^z_k \sigma^z_l \sigma^z_m,
    \end{split}
\end{equation}
a $p$-spin-glass containing up to five-body terms. The Hamiltonian in \cref{eq:protein} represents a sum of three Hamiltonians, $H_f = H_\text{gc} + H_\text{ch} + H_\text{in}$. The geometrical constraint Hamiltonian, $H_\text{gc}$, ensures that a part of the non-physical conformations where the protein folds back into itself are left out of the feasible configuration space (see \cref{app:hardness}). Since the $H_\text{gc}$ is constructed through penalty terms, an appropriate Lagrange multiplier, $\lambda_\text{gc}$, is set to penalize unwanted configurations, and we use $\lambda_\text{gc}=10$. The chirality constraint Hamiltonian $H_\text{ch}$ is included in the formulation if side chains are present, enforcing a correct stereochemistry. In our study, side chains are not considered, hence, this contribution vanishes. Finally, the interaction Hamiltonian, $H_\text{in}$, encodes the interactions among nearest-neighbor beads, where the energy $\epsilon_{ij}$ is added for each nearest-neighbor contact between the pair of non-consecutive beads $(i,j)$~\cite{miyazawa1996residue}. This Hamiltonian term also penalizes the remaining non-physical conformations that are not handled by $H_\text{gc}$ (with the respective Lagrange multiplier also set to 10). For our study, we tackle the following peptides:
\begin{itemize}
    \item Chignolin (\texttt{GYDPETGTWG})~\cite{honda2004residue,chignolin}, a synthetic peptide that folds into a stable $\beta$-hairpin structure, a common motif in real proteins, thus widely used to validate theories. We set $\theta_\text{cutoff}=0.006$ for the soft and $\theta_\text{cutoff}=0.013$ for the hard pruning.
    \item Head activator neuropeptide (\texttt{QPPGGSKVILF})~\cite{bodenmuller1981conserved,morphneuro}, which promotes human neural cell proliferation and differentiation, potentially contributing to regenerative processes in the nervous system. We set $\theta_\text{cutoff}=0.0052$ for the soft and $\theta_\text{cutoff}=0.011$ for the hard pruning.
    \item The immunoglobulin kappa joining 1 gene segment (\texttt{WTFGQGTKVEIK})~\cite{jackson2012divergent,igjk1}, responsible to generate diverse antibodies by joining with other segments during B cell development. We set $\theta_\text{cutoff}=0.005$ for the soft and $\theta_\text{cutoff}=0.013$ for the hard pruning.
\end{itemize}
We used the one-letter code for amino acids, where the letters considered stand for aspartic-acid (\texttt{D}), glutamic-acid (\texttt{E}), phenylalanine (\texttt{F}), glycine (\texttt{G}), isoleucine (\texttt{I}), lysine (\texttt{K}), leucine (\texttt{L}), proline (\texttt{P}), glutamine (\texttt{Q}), serine (\texttt{S}), threonine (\texttt{T}), valine (\texttt{V}), tryptophan (\texttt{W}), and tyrosine (\texttt{Y}). Once the instances are generated, to obtain their exact ground states we use a brute force method.
\begin{table*}[!t]
    \caption{Performance of the BF-DCQO algorithm run on IonQ's Forte and Forte Enterprise quantum hardware for the protein folding, MAX 4-SAT and spin-glass instances tested. The number of Pauli terms is shown for each problem instance. The table compares the best solutions obtained from the QPU execution after 10 iterations, with and without post-processing, with the optimal values determined classically. Bold entries indicate that the optimal solution was found. The number of two-qubit entangling gates, shown as (minimum, maximum) pairs over the 10 iterations for each instance, reflects final counts after circuit compilation.}\label{tab:res}
    \begin{ruledtabular}\begin{tabular}{llrrrrrrr} 
        \multicolumn{1}{c}{Problem} & \multicolumn{1}{c}{Instance} & \multicolumn{1}{c}{Qubits} & \multicolumn{1}{c}{\makecell{Pauli\\terms}} & \multicolumn{1}{c}{$ZZ$ gates (soft)} & \multicolumn{1}{c}{$ZZ$ gates (hard)}  & \multicolumn{1}{c}{\makecell{Best\\solution}} & \multicolumn{1}{c}{\makecell{Best\\solution (PP)}} & \multicolumn{1}{c}{\makecell{Optimal\\solution}} \\ \midrule
        \multirow{3}{*}{\makecell[l]{Protein\\folding}} & \texttt{GYDPETGTWG} [\cref{fig:proteins_forte}(a)] & 22 &  589 & $(630, 646)$ & $(269, 292)$ & $-1.694$ & $\textbf{-2.549}$ & $-2.549$ \\
         & \texttt{QPPGGSKVILF} [\cref{fig:proteins_forte}(b)] & 27 & 901 & $(476, 501)$ & $(349, 350)$ & $-1.886$ & $\textbf{-3.886}$ & $-3.886$ \\
         & \texttt{WTFGQGTKVEIK} [\cref{fig:proteins_forte}(c)] & 33 & 1371 & $(593, 601)$ & $(334, 335)$ & $-2.694$ & $\textbf{-4.328}$ & $-4.328$ \\
        \midrule 
        \multirow{12}{*}{\makecell{MAX 4-SAT}} & 24-1 [\cref{fig:max-4sat}(a)-(b)] & 24 & 1124  & $(215, 747)$ & $(78, 216)$ & \textbf{232} & --- & 232\\
         & 24-2 & 24 & 1157  & $(186, 686)$ & --- & \textbf{231} & --- & 231 \\
         & 24-3 & 24 &  1183 & $(204, 676)$  & --- & \textbf{232} & --- & 232 \\
        \cmidrule{2-9} 
         & 28-1 [\cref{fig:max-4sat}(c)-(d)] & 28 & 1442 & $(189, 698)$ & $(76, 192)$ & 270 & $\textbf{271}$ & 271 \\
         & 28-2 & 28 & 1462 & $(194, 855)$ & --- & 270 & \textbf{271} & 271\\
         & 28-3 & 28 & 1448 & $(212, 716)$ & --- & \textbf{270} & --- & 270 \\
        \cmidrule{2-9}
         & 32-1 [\cref{fig:max-4sat}(e)-(f)] & 32 & 1742 & $(214, 833)$ & $(69, 214)$ & \textbf{310} & --- & 310 \\
         & 32-2 & 32 & 1751 & $(95, 825)$ & --- & \textbf{310} & --- & 310 \\
         & 32-3 & 32 & 1771 & $(235,809)$ & --- & 309 & \textbf{310} & 310 \\
        \cmidrule{2-9} 
         & 36-1 [\cref{fig:max-4sat}(g)-(h)] & 36 & 2084 & $(201, 913)$ & $(75, 202)$ & \textbf{348} & --- & 348 \\
         & 36-2 & 36 & 2093 & $(255, 894)$ & --- & 347 & \textbf{349} & 349 \\
         & 36-3 & 36 & 2076 & $(281, 874)$ & --- & 346 & \textbf{348} & 348 \\
        \midrule
        \multirow{3}{*}{\makecell{Spin-glass}} & 1 [\cref{fig:sk_forteE}(a)] & 36 & 666 & $(310, 364)$ & $(86, 118)$ & $\textbf{-11.405}$ & --- & $-11.405$ \\
         & 2 [\cref{fig:sk_forteE}(b)] & 36 & 666 & $(294, 374)$ & $(62, 110)$ & $-7.365$ & $\textbf{-7.499}$ & $-7.499$ \\
         & 3 [\cref{fig:sk_forteE}(c)] & 36 & 666 & $(404, 684)$ & $(12, 150)$ & $\textbf{-9.966}$ & --- & $-9.966$ \\
    \end{tabular}\end{ruledtabular}
\end{table*}%
\begin{figure*}[!t]
  \centering
  \includegraphics[width=\linewidth]{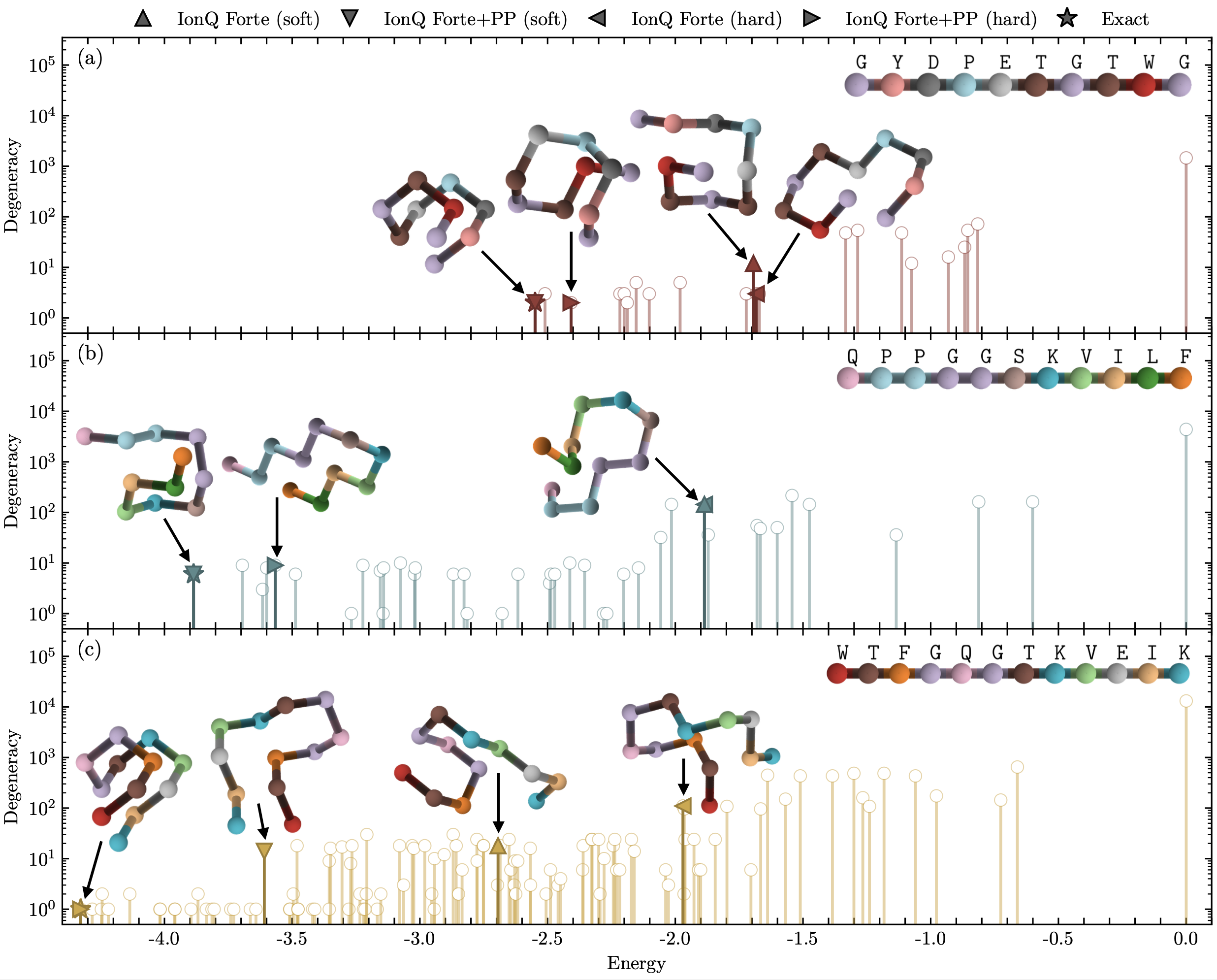}%
\caption{Degeneracy of the lowest-energy valued spectrum for the three proteins studied (in $RT$ units~\cite{miyazawa1996residue}), along with the lowest-energy values obtained after $10$ iterations of the BF-DCQO algorithm executed on quantum hardware using both soft and hard pruning methods and before and after post-processing (triangular markers). (a) \texttt{GYDPETGTWG} (22 qubits, maroon), (b) \texttt{QPPGGSKVILF} (27 qubits, teal) and (c) \texttt{WTFGQGTKVEIK} (33 qubits, gold). The exact ground state is included for comparison (stars), which were obtained using a brute force method. Their corresponding three-dimensional configurations are shown for completeness.
}
\label{fig:proteins_forte}
\end{figure*}%

\subsection{MAX 4-SAT}
\noindent
The maximum $k$-satisfiability (MAX $k$-SAT) problem, a challenge with significant industrial relevance, involves finding an assignment of Boolean variables that maximizes the sum of all the $k$-variable clauses met~\cite{Battiti2009}. If we let $k=4$, this problem reads as
\begin{equation}
    C^\text{M4S}(l) = \bigwedge_{ijkm} l_i\vee l_j \vee l_k \vee l_m,
\end{equation}
with $l_i$ a literal representing a propositional variable, either $u_i$ or its negation $\bar{u}_i$. The problem can be mapped into binary variables by performing the substitution $u_i\mapsto 1-x_i$ and $\bar{u}_i\mapsto x_i$ with $x_i\in\{0,1\}$. To address this problem with quantum computers, we perform an additional map turning the binary variables into Ising variables. Setting $b_i=1$ if $l_i$ is negated and $0$ otherwise, the problem Hamiltonian reads as
\begin{equation}\label{eq:m4s}
\begin{split}
    H_f &= \sum_{c\in C^\text{M4S}} \bigg[ 1 - \frac{1}{16} \prod_{i}[I+(-1)^{b_i}\sigma^z_{i}] \bigg] \\
    &= |C^\text{M4S}| - \frac{1}{16} \sum_{c\in C^\text{M4S}} \prod_{i}[I+(-1)^{b_i}\sigma^z_{i}],
\end{split}
\end{equation}
whose largest eigenvalue corresponds to the maximum number of clauses satisfied. Each summand returns 1 if the clause $c$ is satisfied and 0 otherwise. Since this is a maximization problem, it can be converted into a minimization one by simply negating the Hamiltonian $H^\text{M4S}_f \mapsto -H^\text{M4S}_f$.

The MAX 4-SAT problem results in a highly non-local Hamiltonian, making it challenging to implement on quantum hardware with limited connectiviy. While recent proposals have solved MAX $k$-SAT using superconducting qubits with a purely quantum HUBO approach but constrained to a linear topology~\cite{romero2024bias}, trapped-ions through variational algorithms~\cite{pelofske2023highround}, and quantum annealers after a HUBO-to-QUBO conversion~\cite{zielinski2024solving}, our approach is expected to achieve better performance on more intricate instances by combining the all-to-all connectivity present in IonQ hardware with the BF-DCQO algorithm.

In computer science, computational phase transitions occur at critical points where algorithms require an increasing amount of computational resources, becoming less tractable~\cite{Philathong_2021}. For $k$-SAT, they mainly depend on its clause-to-variable ratio, whose critical value for random 4-SAT is around $\alpha_c=9.7$~\cite{gent1994sat,kirkpatrick1994critical}. Accordingly, we generate three random 4-SAT instances of $N\in\{24,28,32,36\}$ qubits with a number of clauses equal to $\lfloor N\alpha_c\rfloor$. Additionally, for all the instances considered, we set $\theta_\text{cutoff}=0.03$ for the soft and $\theta_\text{cutoff}=0.035$ for the hard pruning. We make use of the \texttt{PySAT} library to obtain a classical reference solution~\cite{ignatiev2024towards, pysat}.

\subsection{All-to-all spin-glass}

\noindent
The Sherrington-Kirkpatrick (SK) spin-glass model \cite{sherrington1975solvable} is a foundational mean-field model in statistical physics, describing a system of interacting spin variables where every spin interacts with every other spin with specified coupling strengths. The Hamiltonian for a problem of $N$ spins can be described as 
\begin{equation}
    H_f = \sum_{i=1}^N h_i\sigma^z_i + \sum_{i<j} J_{ij}\sigma^z_i\sigma^z_j.
\end{equation}

This all-to-all interaction, combined with the quenched disorder (fixed random couplings) and frustration (conflicting interactions that prevent all interactions from being simultaneously satisfied), leads to a highly complex and rugged energy landscape with a vast number of local minima. This characteristic makes the SK model a prototype for understanding disordered systems and their complex energy landscapes. Its relevance to combinatorial optimization problems stems from this very complexity: many hard combinatorial optimization problems, such as finding the maximum cut in a graph or solving satisfiability problems, can be mapped onto finding the ground state (lowest energy configuration) of a spin-glass Hamiltonian, including the SK model. The challenge of navigating the highly non-convex energy landscape of the SK model, with its many local minima, mirrors the computational difficulty of finding optimal solutions in NP-hard combinatorial optimization problems, making the SK model a critical benchmark and a source of insight for developing and testing new optimization algorithms, both classical and quantum.

Here, we study three problem instances of the SK model with 36 spins and random coupling strengths drawn from two separate Gaussian random distributions for the $h_i$ (mean=$0.42$, variance=$0.14$) and $J_{ij}$ (mean=$0.045$, variance=$0.03$) couplings in the Hamiltonian. Moreover, we set $\theta_\text{cutoff}=0.05$ for the soft and $\theta_\text{cutoff}=0.1$ hard pruning. We use Gurobi to obtain a classical reference solution~\cite{gurobi}.
\begin{figure*}[!t]
  \centering
  \includegraphics[width=\linewidth]{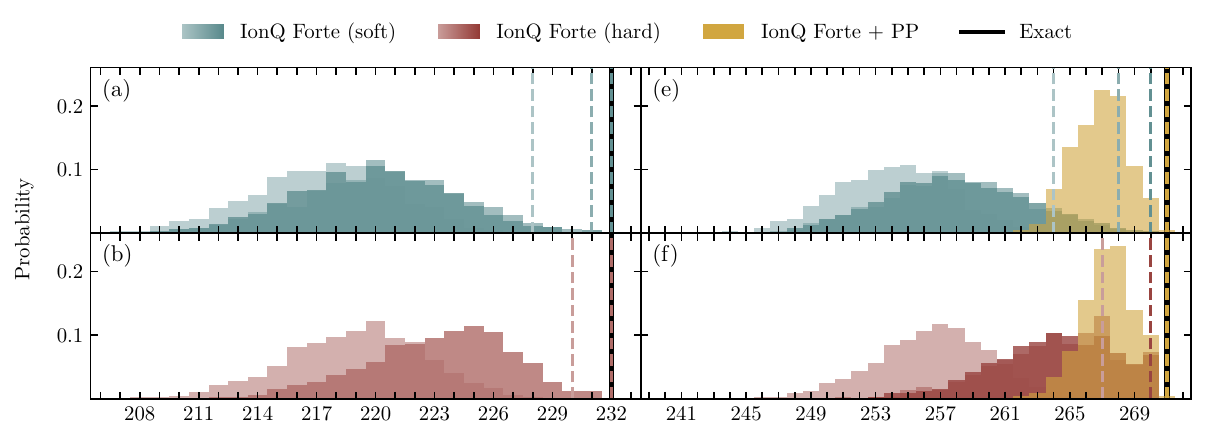}\\\vspace{-3mm}%
  \includegraphics[width=\linewidth]{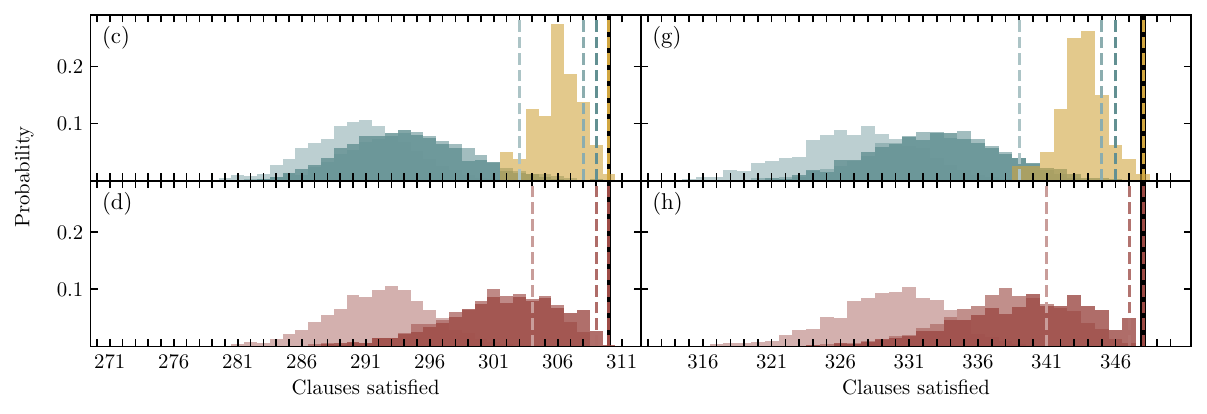}%
\caption{Hardware results of the MAX 4-SAT instances using soft and hard pruning (teal and maroon, respectively), with increasing BF-DCQO iterations going from lighter to darker colors. (a),(b) Instance 24-1. (e),(f) Instance 28-1. (c),(d) Instance 32-1. (g),(h) Instance 36-1. For those instances where optimal solutions were not reached after 10 BF-DCQO iterations, post-processing is applied on top of hardware results (gold), whose corresponding probabilities are divided by six for visualization purposes. Dashed lines indicate the maximum values for their corresponding distributions. Exact solution (black) was obtained using \texttt{PySAT}~\cite{ignatiev2024towards, pysat}. }
\label{fig:max-4sat}
\end{figure*}%
\begin{figure*}[!t]
  \centering
  \includegraphics[width=1\linewidth]{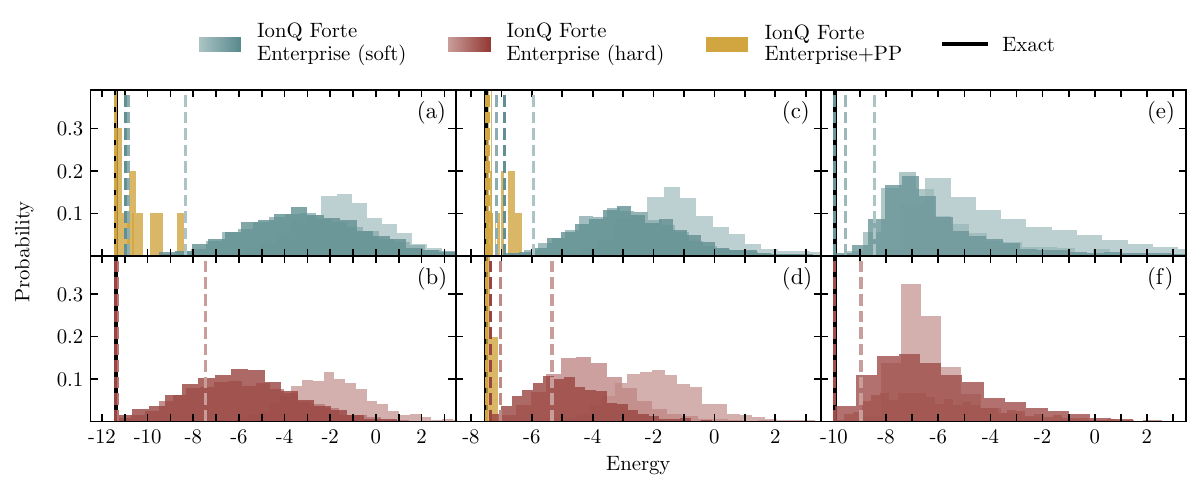}%
\caption{Spin-glass Forte Enterprise results using soft and hard pruning (teal and maroon, respectively), with increasing BF-DCQO iterations going from lighter to darker colors. For those instances where optimal solutions were not reached after 10 BF-DCQO iterations, post-processing is applied on top of hardware results (gold). Dashed lines indicate the minimum values for their corresponding distributions. Exact solution (black) is obtained using Gurobi~\cite{gurobi}. (a),(b) spin-glass instance 1. (c),(d) spin-glass instance 2. (e),(f) spin-glass instance 3. 
}
\label{fig:sk_forteE}
\end{figure*}%

\section{Results and discussion}\label{sec:results}

\noindent
In this section, we briefly introduce the specifications of the IonQ Forte devices, and we present the different problems to be solved using these architectures. To showcase the efficiency of our proposal, each iteration of BF-DCQO was run with $2000$ shots across all instances. Furthermore, if the optimal solution was not directly obtained from hardware, a minimal local search algorithm is applied on top of our best sampled results, which can be seen as a post-processing technique (referred to as PP) by means of a simulated annealing method with zero-temperature Metropolis-Hastings criterion~\cite{metropolis1953equation,hastings1970monte}. In particular, we take the 5\% best samples and apply up to three sweeps. 

\subsection{Trapped-ion quantum hardware}
\label{subsec:IonQ}
\noindent
IonQ's quantum processing units (QPUs), specifically the IonQ Forte and IonQ Forte Enterprise systems~\cite{Chen2024-co}, house 36 qubits each. These qubits are realized using trapped Ytterbium ions ($^{171}$Yb$^+$), with quantum information encoded in their ground state hyperfine levels. The ions are precisely sourced through laser ablation and selective ionization, then loaded into a compact, integrated vacuum package containing a surface linear Paul trap.

Qubit states are manipulated using $\SI{355}{nm}$ laser pulses, which drive two-photon Raman transitions. This enables a universal set of operations, including arbitrary single-qubit rotations and two-qubit $ZZ$ entangling gates (further details in \cref{app:circuit}). At the time of execution, the median direct randomized benchmarking (DRB) fidelities for these entangling gates were $99.3\%$ on Forte and $99.5\%$ on Forte Enterprise and around $99.98\%$ for the single qubit rotation gates. The gate durations were around 950 microseconds for the two-qubit $ZZ$ gates and around 130 microseconds for single-qubit gates.

Both QPU types integrate acousto-optic deflectors (AODs) for independent laser beam steering to each ion, significantly reducing beam alignment errors~\cite{Kim:2008ApOpt,Pogorelov:2021PRXQ}. This advanced optical and robust control system, which automates calibration and optimizes gate execution, has allowed IonQ to build larger qubit registers with consistently high gate fidelities. Ultimately, IonQ Forte-generation devices represent a scalable, high-fidelity platform for quantum information processing, marking a significant advancement in trapped-ion QPU technology.

\subsection{Quantum hardware results}

\noindent
The experimental results and performance of the BF-DCQO algorithm on the trapped-ion quantum hardware are summarized in \cref{tab:res}. This table contains the best solutions obtained for all instances across the three classes of problems studied: protein folding, MAX 4-SAT, and two-body spin-glass.

For protein folding, BF-DCQO (combined with PP) found optimal solutions for all the amino acid sequences tested, as well as high-quality, sub-optimal solutions (\cref{tab:res}). For the suboptimal results, the energy difference between the BF-DCQO solutions and the true ground state was between $10^{-1}$ and $10^0$. This difference is negligible compared to the problem's total spectral range of approximately $10^5$ (see \cref{app:hardness}). Moreover, the spectrum contains a large density of highly degenerate energy levels, including, in some cases, a degenerate ground state. 

Figure~\ref{fig:proteins_forte} illustrates the high degeneracy of the low-energy spectrum and compares the protein structures obtained by our algorithm to the optimal configurations. This figure clearly highlights the complexity of this problem, namely, to reach the ground state, a solver must navigate a dense landscape of degenerate excited states separated by small energy gaps. The high-quality approximate solutions obtained on hardware were refined to the true optimal solutions for all the amino acid chains by applying minimal post-processing (shown as PP in \cref{tab:res,fig:proteins_forte}). Notably, these instances represent the largest protein folding instances successfully solved on trapped-ion hardware to date, highlighting the value of combining our adapted protein folding solver based on BF-DCQO with trapped-ion hardware.

In the case of the MAX 4-SAT instances, the algorithm found optimal solutions for all of the instances (see \cref{tab:res}). Notably, for each of the $24$, $28$, $32$, and $36$ qubit problem sizes, at least one optimal solution was identified. In cases where optimality was not reached, the solutions were marginally suboptimal, differing by only a single clause (or two for the 36-qubit instances). For this problem class, three distinct instances were tested for each size; while soft pruning was applied to all three, hard pruning was used on only a single instance per problem size. These findings show that hard pruning consistently outperformed soft pruning, a result likely attributable to hardware-induced noise; soft-pruned circuits have greater depth and are thus more susceptible to error propagation. This was particularly noticeable in the 36-qubit case, where the hard-pruned circuit was the only one that yielded an optimal solution. We present the results obtained on hardware in \cref{fig:max-4sat}. 

For the two-body spin-glass problem class, BF-DCQO successfully identified the exact ground state energies of two of the three 36-qubit all-to-all connected instances in fewer than ten iterations (see \cref{tab:res}). Similarly to the MAX 4-SAT instances, hard-pruned circuits proved to be more efficient in quantum hardware. In the single instance where the ground state was not found, the best solution was of high quality. We present the histograms associated with this problem in \cref{fig:sk_forteE}.

It is worth mentioning that the practical implementation of this algorithm was contingent on circuit pruning. In \cref{tab:specs}, we show the significant reduction in gate count from the original to the pruned circuits, confirming that the full versions would have circuit depths too large to be executed on current-generation quantum hardware.

As a concluding remark and as expected from previous works~\cite{romero2024bias}, in our experiments, the BF-DCQO algorithm required a number of resources that reduces with the number of iterations, despite the outcomes obtained being more optimal. This feature makes our algorithm remarkably suitable for current and near-term hardware, combining implementability with efficacy.

\section{Conclusion}\label{sec:conclusion}

\noindent
We introduce a quantum protein folding solver based on the BF-DCQO algorithm and adapted to current trapped-ion architectures. We obtain optimal 
solutions with minimal resources on the largest-to-date protein-folding instances solved with trapped ions. Featuring high gate fidelities and all-to-all connectivity, IonQ platforms provide a natural setup to implement instances with several terms and long-range interactions, commonly present in optimization problems but challenging to realize on up-to-date quantum devices. To assess its performance, we solve protein folding instances going up to 33 qubits, obtaining both optimal and near-optimal solutions. The flexibility of our method is showcased by also solving random all-to-all connected spin-glass problems and MAX 4-SAT instances up to 36 qubits. When optimal solutions are not directly obtained, we additionally apply a greedy local search algorithm to get better solutions by mitigating potential bit-flip and measurement errors, which highlights the favorable interplay between classical and quantum algorithms. Led by our results, our proposal offers a non-variational approach that efficiently solves protein folding with high accuracy. More optimal solutions are obtained despite the quantum resource requirement reducing with subsequent BF-DCQO iterations. In this manner, BF-DCQO emerges as a suitable algorithm for solving large-scale protein folding instances on current and next-generation trapped-ion quantum processors.

\begin{acknowledgments}\noindent
    We thank Michael Falkenthal, Michael Wurster, and Sebastian Wagner for their assistance in setting up the services on Kipu Quantum's PLANQK platform required to run the experiments. We also thank Pranav Chandarana, Gaurav Dev, Shubham Kumar, and Anton Simen for useful discussions. Miguel Angel Lopez-Ruiz, Claudio Girotto, Masako Yamada, Panagiotis Kl. Barkoutsos, Ananth Kaushik, and Martin Roetteler are employees and equity holders of IonQ Inc.
\end{acknowledgments}

\noindent\textbf{Data availability statement.} The data used in the current study are available upon reasonable request from the corresponding authors.

\appendix

\section{Circuit decomposition}\label{app:circuit}

\begin{figure*}[!t]
  \centering
  \includegraphics[width=\linewidth]{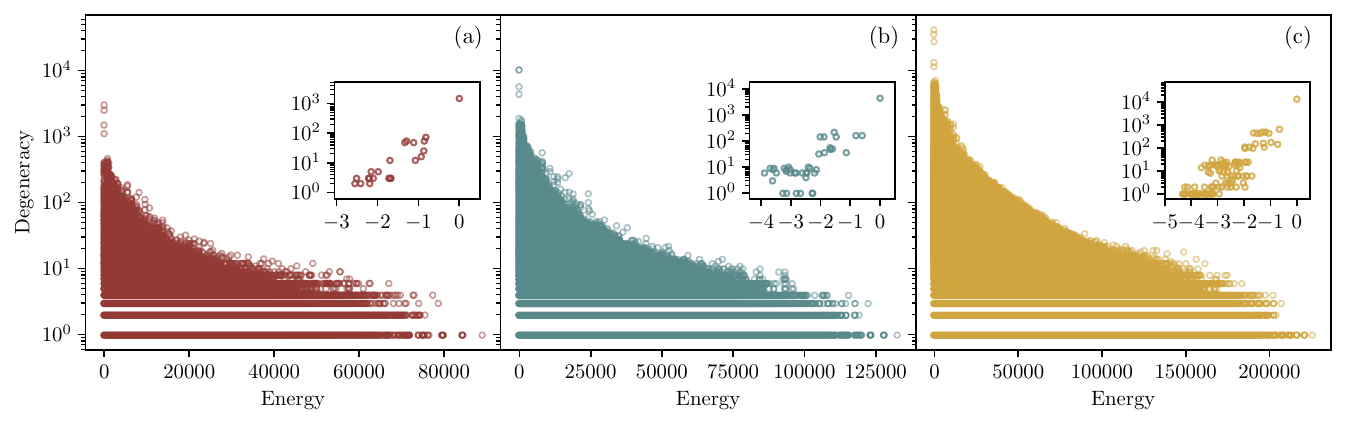}%
\caption{Degeneracy of the full energy spectrum (in $RT$ units~\cite{miyazawa1996residue}) of the (a) \texttt{GYDPETGTWG} (22 qubits, maroon), (b) \texttt{QPPGGSKVILF} (27 qubits, teal), and (c) \texttt{WTFGQGTKVEIK} (33 qubits, gold) protein instances, respectively. Zoomed insets show the degeneracy of the lowest energy-valued configurations, region where the most optimal foldings are located.}
\label{fig:protein_22_spectrum}
\end{figure*}%

\noindent
A key aspect to account for when preparing and running quantum circuits on hardware is to transpile the required quantum operations according to the corresponding native gate sets provided by the platform. These typically consist of a universal gate set containing several one-qubit gates and a single two-qubit entangling gate. In our experiments, we use the IonQ Forte and Forte Enterprise hardware, whose native gate set is composed of the one-qubit gates
\begin{equation}
\begin{split}
    \text{VirtualZ}(\phi)&=\exp\left(-i\frac{\phi}{2}\sigma^z\right)=\begin{bmatrix} e^{-i\phi/2} & 0 \\ 0 & e^{i\phi/2}\end{bmatrix}, \\
    \text{GPi}(\phi) &= \begin{bmatrix} 0 & e^{-i\phi} \\ e^{i\phi} & 0 \end{bmatrix}, \\
    \text{GPi2}(\phi) &= \frac{1}{\sqrt{2}} \begin{bmatrix} 1 & -ie^{-i\phi} \\ -ie^{i\phi} & 1 \end{bmatrix},
\end{split}
\end{equation}
and the entangling gate
\begin{equation}
\begin{split}
    ZZ(\phi) &= \exp\left(-i\frac{\phi}{2}\sigma^z_0\sigma^z_1\right)\\
    &=\begin{bmatrix}
        e^{-i\phi/2}&0&0&0\\
        0&e^{i\phi/2}&0&0\\
        0&0&e^{i\phi/2}&0\\
        0&0&0&e^{-i\phi/2}
    \end{bmatrix}.
\end{split}
\end{equation}

\section{About the hardness of protein folding}\label{app:hardness}

\noindent
One of the inherent features of protein folding instances built by Ref.~\cite{robert2021resource} corresponds to their highly degenerate spectrum, the small gap between the ground and first excited states, and the large difference between the ground and most excited states, mainly caused by the inclusion of penalty terms in the formulation. In \cref{fig:protein_22_spectrum}, we plot the degeneracy of the spectrum of the protein instances considered, seeing how the lowest energy-valued states (and often even the ground state) are highly degenerated, which can hinder the convergence of optimization algorithms. The incorporation of the CD contribution into the adiabatic evolution translates into an increase of the spectral gap during the evolution~\cite{hegade2022digitized}, which helps to reduce the excitations, resulting in a better convergence and an increased success probability.

Apart from that, we can compute how the number of feasible and unfeasible solutions scale with the number of amino acids $N_a$. Since no side chains are considered in our study, chirality constraints are omitted; therefore, we elucidate the scaling by looking at the different possible configurations under the encoding of Ref.~\cite{robert2021resource}.

The total set of qubits is partitioned into configuration and interaction subsets.
The configuration qubits encode the sequence of turns in the protein chain, 
where each turn $t_i$ is encoded by a pair of qubits, namely, $t_i=q_{2i-1}q_{2i}$. For a protein with $N_a$ amino acids, this scales as $2(N_a-1)$ configuration qubits. At each turn, the sublattices $\mathcal{A}$ and $\mathcal{B}$ are interchanged, with even (odd) turns corresponding to $\mathcal{A}$ ($\mathcal{B}$) (in a tetrahedral lattice, there are two sets of symmetrically distinct points (see Ref.~\cite{robert2021resource} for an in-depth construction)). Due to spatial symmetry, without loss of generality, we can fix the first two turns as $t_1=\bar{1}$ and $t_2=0$ as well as $q_6=1$, where the remaining $2(N_a-3)-1$ qubits have to be optimized. Therefore, the conformational structure of the protein is represented as
\begin{equation}
    [01][00][q_51][q_7q_8]\cdots[q_{2(N_a-1)-1}q_{2(N_a-1)}].
\end{equation}
If two consecutive sets lay within the same axis, e.g., $\bar{0}0$, then the protein folds back into itself, returning an unfeasible conformation. Nonetheless, farther neighbors can also fold back into previously set positions. We can derive an upper bound in the number of feasible configurations $n_f(N_a)$ by computing between nearest-neighbors (NN) the number of feasible configurations that does not share axis consecutively $n^\text{NN}_f(N_a)$. Therefore, $n_f(N_a)<n^\text{NN}_f(N_a)=2\cdot3^{N_a-4}\sim\mathcal{O}[3^{N_a}]$ configurations are feasible while $n_u(N_a)=2\cdot4^{N_a-4}-n_f(N_a)\sim\mathcal{O}[4^{N_a}]$ are not. This indicates that the number of feasible configurations decreases exponentially with increasing $N_a$.

\section{Pruning and implementability}
\label{app:pruning}

\noindent
In this Appendix, we explicitly show the effect of pruning on implementability on noisy quantum hardware. Specifically, we track the number of entangling gates for the particular choice of pruning, which usually feature higher error rates and slower running times when compared to one-qubit gates. For the soft-pruning case, we keep the number of entangling gates close to a thousand, whereas for hard-pruning the number of gates is in the hundreds, e.g., see \cref{tab:specs} for the MAX 4-SAT problems. 
\begin{table}[!h]
    \caption{In parentheses, minimum and maximum values for the MAX 4-SAT instances. For the the last column, in the left (right) tuple we apply a soft (hard) pruning.}\label{tab:specs}
    \begin{ruledtabular}\begin{tabular}{rrr}
       \multicolumn{1}{c}{Qubits} & \multicolumn{1}{c}{$ZZ$ gates (no pruning)} & \multicolumn{1}{c}{$ZZ$ gates (pruning)} \\ \midrule
       24 & $(7390,7578)$ & $(644,762)$ / $(148,216)$ \\
       28 & $(9106, 9324)$ & $(666,826)$ / $(166, 238)$ \\
       32 & $(10938,11100)$ & $(778,936)$ / $(192,274)$ \\
       36 & $(12802,13028)$ & $(872, 1020)$ / $(208,270)$
    \end{tabular}\end{ruledtabular}
\end{table}%

\bibliography{bibfile}

\begin{thebibliography}{95}%
\makeatletter
\providecommand \@ifxundefined [1]{%
 \@ifx{#1\undefined}
}%
\providecommand \@ifnum [1]{%
 \ifnum #1\expandafter \@firstoftwo
 \else \expandafter \@secondoftwo
 \fi
}%
\providecommand \@ifx [1]{%
 \ifx #1\expandafter \@firstoftwo
 \else \expandafter \@secondoftwo
 \fi
}%
\providecommand \natexlab [1]{#1}%
\providecommand \enquote  [1]{``#1''}%
\providecommand \bibnamefont  [1]{#1}%
\providecommand \bibfnamefont [1]{#1}%
\providecommand \citenamefont [1]{#1}%
\providecommand \href@noop [0]{\@secondoftwo}%
\providecommand \href [0]{\begingroup \@sanitize@url \@href}%
\providecommand \@href[1]{\@@startlink{#1}\@@href}%
\providecommand \@@href[1]{\endgroup#1\@@endlink}%
\providecommand \@sanitize@url [0]{\catcode `\\12\catcode `\$12\catcode `\&12\catcode `\#12\catcode `\^12\catcode `\_12\catcode `\%12\relax}%
\providecommand \@@startlink[1]{}%
\providecommand \@@endlink[0]{}%
\providecommand \url  [0]{\begingroup\@sanitize@url \@url }%
\providecommand \@url [1]{\endgroup\@href {#1}{\urlprefix }}%
\providecommand \urlprefix  [0]{URL }%
\providecommand \Eprint [0]{\href }%
\providecommand \doibase [0]{https://doi.org/}%
\providecommand \selectlanguage [0]{\@gobble}%
\providecommand \bibinfo  [0]{\@secondoftwo}%
\providecommand \bibfield  [0]{\@secondoftwo}%
\providecommand \translation [1]{[#1]}%
\providecommand \BibitemOpen [0]{}%
\providecommand \bibitemStop [0]{}%
\providecommand \bibitemNoStop [0]{.\EOS\space}%
\providecommand \EOS [0]{\spacefactor3000\relax}%
\providecommand \BibitemShut  [1]{\csname bibitem#1\endcsname}%
\let\auto@bib@innerbib\@empty
\bibitem [{\citenamefont {Kirkpatrick}\ \emph {et~al.}(1983)\citenamefont {Kirkpatrick}, \citenamefont {Gelatt},\ and\ \citenamefont {Vecchi}}]{kirkpatrick1983optimization}%
  \BibitemOpen
  \bibfield  {author} {\bibinfo {author} {\bibfnamefont {S.}~\bibnamefont {Kirkpatrick}}, \bibinfo {author} {\bibfnamefont {C.~D.}\ \bibnamefont {Gelatt}},\ and\ \bibinfo {author} {\bibfnamefont {M.~P.}\ \bibnamefont {Vecchi}},\ }\href {https://doi.org/10.1126/science.220.4598.671} {\bibfield  {journal} {\bibinfo  {journal} {Science}\ }\textbf {\bibinfo {volume} {220}},\ \bibinfo {pages} {671} (\bibinfo {year} {1983})}\BibitemShut {NoStop}%
\bibitem [{\citenamefont {Glover}\ \emph {et~al.}(1993)\citenamefont {Glover}, \citenamefont {Taillard},\ and\ \citenamefont {de~Werra}}]{tabu}%
  \BibitemOpen
  \bibfield  {author} {\bibinfo {author} {\bibfnamefont {F.}~\bibnamefont {Glover}}, \bibinfo {author} {\bibfnamefont {E.}~\bibnamefont {Taillard}},\ and\ \bibinfo {author} {\bibfnamefont {D.}~\bibnamefont {de~Werra}},\ }\href {https://doi.org/10.1007/BF02078647} {\bibfield  {journal} {\bibinfo  {journal} {Annals of Operations Research}\ }\textbf {\bibinfo {volume} {41}},\ \bibinfo {pages} {3} (\bibinfo {year} {1993})}\BibitemShut {NoStop}%
\bibitem [{\citenamefont {Cplex}(2009)}]{cplex}%
  \BibitemOpen
  \bibfield  {author} {\bibinfo {author} {\bibfnamefont {I.~I.}\ \bibnamefont {Cplex}},\ }\href@noop {} {\bibfield  {journal} {\bibinfo  {journal} {International Business Machines Corporation}\ }\textbf {\bibinfo {volume} {46}},\ \bibinfo {pages} {157} (\bibinfo {year} {2009})}\BibitemShut {NoStop}%
\bibitem [{\citenamefont {{Gurobi Optimization, LLC}}(2023)}]{gurobi}%
  \BibitemOpen
  \bibfield  {author} {\bibinfo {author} {\bibnamefont {{Gurobi Optimization, LLC}}},\ }\href {https://www.gurobi.com} {\bibinfo {title} {{Gurobi Optimizer Reference Manual}}} (\bibinfo {year} {2023})\BibitemShut {NoStop}%
\bibitem [{\citenamefont {Bertsekas}\ \emph {et~al.}(2003)\citenamefont {Bertsekas}, \citenamefont {Nedic},\ and\ \citenamefont {Ozdaglar}}]{bertsekas2003convex}%
  \BibitemOpen
  \bibfield  {author} {\bibinfo {author} {\bibfnamefont {D.~P.}\ \bibnamefont {Bertsekas}}, \bibinfo {author} {\bibfnamefont {A.}~\bibnamefont {Nedic}},\ and\ \bibinfo {author} {\bibfnamefont {A.~E.}\ \bibnamefont {Ozdaglar}},\ }\href {https://www.athenasc.com/convexity.html} {\emph {\bibinfo {title} {Convex Analysis and Optimization}}}\ (\bibinfo  {publisher} {Athena Scientific},\ \bibinfo {address} {Belmont, MA},\ \bibinfo {year} {2003})\BibitemShut {NoStop}%
\bibitem [{\citenamefont {Floudas}\ \emph {et~al.}(2005)\citenamefont {Floudas}, \citenamefont {Akrotirianakis}, \citenamefont {Caratzoulas}, \citenamefont {Meyer},\ and\ \citenamefont {Kallrath}}]{nonlinearprog}%
  \BibitemOpen
  \bibfield  {author} {\bibinfo {author} {\bibfnamefont {C.~A.}\ \bibnamefont {Floudas}}, \bibinfo {author} {\bibfnamefont {I.~G.}\ \bibnamefont {Akrotirianakis}}, \bibinfo {author} {\bibfnamefont {S.}~\bibnamefont {Caratzoulas}}, \bibinfo {author} {\bibfnamefont {C.~A.}\ \bibnamefont {Meyer}},\ and\ \bibinfo {author} {\bibfnamefont {J.}~\bibnamefont {Kallrath}},\ }\href {https://doi.org/10.1016/j.compchemeng.2005.02.006} {\bibfield  {journal} {\bibinfo  {journal} {Computers \& Chemical Engineering}\ }\textbf {\bibinfo {volume} {29}},\ \bibinfo {pages} {1185} (\bibinfo {year} {2005})}\BibitemShut {NoStop}%
\bibitem [{\citenamefont {Burer}\ and\ \citenamefont {Letchford}(2012)}]{surveynonconvex}%
  \BibitemOpen
  \bibfield  {author} {\bibinfo {author} {\bibfnamefont {S.}~\bibnamefont {Burer}}\ and\ \bibinfo {author} {\bibfnamefont {A.~N.}\ \bibnamefont {Letchford}},\ }\href {https://doi.org/10.1016/j.sorms.2012.08.001} {\bibfield  {journal} {\bibinfo  {journal} {Surveys in Operations Research and Management Science}\ }\textbf {\bibinfo {volume} {17}},\ \bibinfo {pages} {97} (\bibinfo {year} {2012})}\BibitemShut {NoStop}%
\bibitem [{\citenamefont {Arjevani}\ \emph {et~al.}(2020)\citenamefont {Arjevani}, \citenamefont {Carmon}, \citenamefont {Duchi}, \citenamefont {Foster}, \citenamefont {Sekhari},\ and\ \citenamefont {Sridharan}}]{arjevani2020nonconvexchallenges}%
  \BibitemOpen
  \bibfield  {author} {\bibinfo {author} {\bibfnamefont {Y.}~\bibnamefont {Arjevani}}, \bibinfo {author} {\bibfnamefont {Y.}~\bibnamefont {Carmon}}, \bibinfo {author} {\bibfnamefont {J.~C.}\ \bibnamefont {Duchi}}, \bibinfo {author} {\bibfnamefont {D.~J.}\ \bibnamefont {Foster}}, \bibinfo {author} {\bibfnamefont {A.}~\bibnamefont {Sekhari}},\ and\ \bibinfo {author} {\bibfnamefont {K.}~\bibnamefont {Sridharan}},\ }in\ \href {https://proceedings.mlr.press/v125/arjevani20a.html} {\emph {\bibinfo {booktitle} {{Proceedings of the 33rd Conference on Learning Theory}}}},\ \bibinfo {series} {Proceedings of Machine Learning Research}, Vol.\ \bibinfo {volume} {125}\ (\bibinfo  {publisher} {PMLR},\ \bibinfo {year} {2020})\ pp.\ \bibinfo {pages} {242--299}\BibitemShut {NoStop}%
\bibitem [{\citenamefont {Danilova}\ \emph {et~al.}(2022)\citenamefont {Danilova}, \citenamefont {Dvurechensky}, \citenamefont {Gasnikov}, \citenamefont {Gorbunov}, \citenamefont {Guminov}, \citenamefont {Kamzolov},\ and\ \citenamefont {Shibaev}}]{danilova2022nonconvexchallenges}%
  \BibitemOpen
  \bibfield  {author} {\bibinfo {author} {\bibfnamefont {M.}~\bibnamefont {Danilova}}, \bibinfo {author} {\bibfnamefont {P.}~\bibnamefont {Dvurechensky}}, \bibinfo {author} {\bibfnamefont {A.}~\bibnamefont {Gasnikov}}, \bibinfo {author} {\bibfnamefont {E.}~\bibnamefont {Gorbunov}}, \bibinfo {author} {\bibfnamefont {S.}~\bibnamefont {Guminov}}, \bibinfo {author} {\bibfnamefont {D.}~\bibnamefont {Kamzolov}},\ and\ \bibinfo {author} {\bibfnamefont {I.}~\bibnamefont {Shibaev}},\ }in\ \href {https://doi.org/10.1007/978-3-031-00832-0_3} {\emph {\bibinfo {booktitle} {{High-Dimensional Optimization and Probability: With a View Towards Data Science}}}},\ \bibinfo {series} {Springer Optimization and Its Applications}, Vol.\ \bibinfo {volume} {191}\ (\bibinfo  {publisher} {Springer},\ \bibinfo {year} {2022})\ pp.\ \bibinfo {pages} {79--163}\BibitemShut {NoStop}%
\bibitem [{\citenamefont {Xu}\ and\ \citenamefont {Liberti}(2024)}]{xu2024relaxationschallenges}%
  \BibitemOpen
  \bibfield  {author} {\bibinfo {author} {\bibfnamefont {L.}~\bibnamefont {Xu}}\ and\ \bibinfo {author} {\bibfnamefont {L.}~\bibnamefont {Liberti}},\ }\href {https://arxiv.org/abs/2405.13447} {\bibinfo {title} {Relaxations for binary polynomial optimization via signed certificates}} (\bibinfo {year} {2024}),\ \Eprint {https://arxiv.org/abs/2405.13447} {arXiv:2405.13447 [math.OC]} \BibitemShut {NoStop}%
\bibitem [{\citenamefont {Lucas}(2014)}]{lucas2014ising}%
  \BibitemOpen
  \bibfield  {author} {\bibinfo {author} {\bibfnamefont {A.}~\bibnamefont {Lucas}},\ }\href {https://doi.org/10.3389/fphy.2014.00005} {\bibfield  {journal} {\bibinfo  {journal} {Frontiers in Physics}\ }\textbf {\bibinfo {volume} {2}},\ \bibinfo {pages} {5} (\bibinfo {year} {2014})}\BibitemShut {NoStop}%
\bibitem [{\citenamefont {Gardner}(1985)}]{gardner1985spin}%
  \BibitemOpen
  \bibfield  {author} {\bibinfo {author} {\bibfnamefont {E.}~\bibnamefont {Gardner}},\ }\href {https://doi.org/https://doi.org/10.1016/0550-3213(85)90374-8} {\bibfield  {journal} {\bibinfo  {journal} {Nuclear Physics B}\ }\textbf {\bibinfo {volume} {257}},\ \bibinfo {pages} {747} (\bibinfo {year} {1985})}\BibitemShut {NoStop}%
\bibitem [{\citenamefont {Pelofske}\ \emph {et~al.}(2023{\natexlab{a}})\citenamefont {Pelofske}, \citenamefont {B{\"a}rtschi},\ and\ \citenamefont {Eidenbenz}}]{pelofske2023quantum}%
  \BibitemOpen
  \bibfield  {author} {\bibinfo {author} {\bibfnamefont {E.}~\bibnamefont {Pelofske}}, \bibinfo {author} {\bibfnamefont {A.}~\bibnamefont {B{\"a}rtschi}},\ and\ \bibinfo {author} {\bibfnamefont {S.}~\bibnamefont {Eidenbenz}},\ }in\ \href {https://doi.org/10.1007/978-3-031-32041-5_13} {\emph {\bibinfo {booktitle} {{High Performance Computing: 38th International Conference, ISC High Performance 2023, Hamburg, Germany, May 21--25, 2023, Proceedings}}}},\ \bibinfo {series} {Lecture Notes in Computer Science}, Vol.\ \bibinfo {volume} {13944}\ (\bibinfo  {publisher} {Springer Nature Switzerland},\ \bibinfo {address} {Cham},\ \bibinfo {year} {2023})\ pp.\ \bibinfo {pages} {240--258}\BibitemShut {NoStop}%
\bibitem [{\citenamefont {Pelofske}\ \emph {et~al.}(2024)\citenamefont {Pelofske}, \citenamefont {Bärtschi},\ and\ \citenamefont {Eidenbenz}}]{pelofske2024short-depth}%
  \BibitemOpen
  \bibfield  {author} {\bibinfo {author} {\bibfnamefont {E.}~\bibnamefont {Pelofske}}, \bibinfo {author} {\bibfnamefont {A.}~\bibnamefont {Bärtschi}},\ and\ \bibinfo {author} {\bibfnamefont {S.}~\bibnamefont {Eidenbenz}},\ }\href {https://doi.org/10.1038/s41534-024-00825-w} {\bibfield  {journal} {\bibinfo  {journal} {npj Quantum Information}\ }\textbf {\bibinfo {volume} {10}},\ \bibinfo {pages} {1–19} (\bibinfo {year} {2024})}\BibitemShut {NoStop}%
\bibitem [{\citenamefont {Albash}\ and\ \citenamefont {Lidar}(2018)}]{albash2018adiabatic}%
  \BibitemOpen
  \bibfield  {author} {\bibinfo {author} {\bibfnamefont {T.}~\bibnamefont {Albash}}\ and\ \bibinfo {author} {\bibfnamefont {D.~A.}\ \bibnamefont {Lidar}},\ }\href {https://journals.aps.org/rmp/abstract/10.1103/RevModPhys.90.015002} {\bibfield  {journal} {\bibinfo  {journal} {Reviews of Modern Physics}\ }\textbf {\bibinfo {volume} {90}},\ \bibinfo {pages} {015002} (\bibinfo {year} {2018})}\BibitemShut {NoStop}%
\bibitem [{\citenamefont {Farhi}\ \emph {et~al.}(2014)\citenamefont {Farhi}, \citenamefont {Goldstone},\ and\ \citenamefont {Gutmann}}]{farhi2014quantumapproximateoptimizationalgorithm}%
  \BibitemOpen
  \bibfield  {author} {\bibinfo {author} {\bibfnamefont {E.}~\bibnamefont {Farhi}}, \bibinfo {author} {\bibfnamefont {J.}~\bibnamefont {Goldstone}},\ and\ \bibinfo {author} {\bibfnamefont {S.}~\bibnamefont {Gutmann}},\ }\href {https://arxiv.org/abs/1411.4028} {\bibinfo {title} {{A Quantum Approximate Optimization Algorithm}}} (\bibinfo {year} {2014}),\ \Eprint {https://arxiv.org/abs/1411.4028} {arXiv:1411.4028 [quant-ph]} \BibitemShut {NoStop}%
\bibitem [{\citenamefont {Barends}\ \emph {et~al.}(2016)\citenamefont {Barends}, \citenamefont {Shabani}, \citenamefont {Lamata}, \citenamefont {Kelly}, \citenamefont {Mezzacapo}, \citenamefont {Heras}, \citenamefont {Babbush}, \citenamefont {Fowler}, \citenamefont {Campbell}, \citenamefont {Chen} \emph {et~al.}}]{barends2016digitized}%
  \BibitemOpen
  \bibfield  {author} {\bibinfo {author} {\bibfnamefont {R.}~\bibnamefont {Barends}}, \bibinfo {author} {\bibfnamefont {A.}~\bibnamefont {Shabani}}, \bibinfo {author} {\bibfnamefont {L.}~\bibnamefont {Lamata}}, \bibinfo {author} {\bibfnamefont {J.}~\bibnamefont {Kelly}}, \bibinfo {author} {\bibfnamefont {A.}~\bibnamefont {Mezzacapo}}, \bibinfo {author} {\bibfnamefont {U.~L.}\ \bibnamefont {Heras}}, \bibinfo {author} {\bibfnamefont {R.}~\bibnamefont {Babbush}}, \bibinfo {author} {\bibfnamefont {A.~G.}\ \bibnamefont {Fowler}}, \bibinfo {author} {\bibfnamefont {B.}~\bibnamefont {Campbell}}, \bibinfo {author} {\bibfnamefont {Y.}~\bibnamefont {Chen}}, \emph {et~al.},\ }\href {https://doi.org/10.1038/nature17658} {\bibfield  {journal} {\bibinfo  {journal} {Nature}\ }\textbf {\bibinfo {volume} {534}},\ \bibinfo {pages} {222–226} (\bibinfo {year} {2016})}\BibitemShut {NoStop}%
\bibitem [{\citenamefont {Abbas}\ \emph {et~al.}(2024)\citenamefont {Abbas}, \citenamefont {Ambainis}, \citenamefont {Augustino}, \citenamefont {Bärtschi}, \citenamefont {Buhrman}, \citenamefont {Coffrin}, \citenamefont {Cortiana}, \citenamefont {Dunjko}, \citenamefont {Egger}, \citenamefont {Elmegreen} \emph {et~al.}}]{abbas2024challenges}%
  \BibitemOpen
  \bibfield  {author} {\bibinfo {author} {\bibfnamefont {A.}~\bibnamefont {Abbas}}, \bibinfo {author} {\bibfnamefont {A.}~\bibnamefont {Ambainis}}, \bibinfo {author} {\bibfnamefont {B.}~\bibnamefont {Augustino}}, \bibinfo {author} {\bibfnamefont {A.}~\bibnamefont {Bärtschi}}, \bibinfo {author} {\bibfnamefont {H.}~\bibnamefont {Buhrman}}, \bibinfo {author} {\bibfnamefont {C.}~\bibnamefont {Coffrin}}, \bibinfo {author} {\bibfnamefont {G.}~\bibnamefont {Cortiana}}, \bibinfo {author} {\bibfnamefont {V.}~\bibnamefont {Dunjko}}, \bibinfo {author} {\bibfnamefont {D.~J.}\ \bibnamefont {Egger}}, \bibinfo {author} {\bibfnamefont {B.~G.}\ \bibnamefont {Elmegreen}}, \emph {et~al.},\ }\href {https://doi.org/10.1038/s42254-024-00770-9} {\bibfield  {journal} {\bibinfo  {journal} {Nature Reviews Physics}\ }\textbf {\bibinfo {volume} {6}},\ \bibinfo {pages} {718–735} (\bibinfo {year} {2024})}\BibitemShut {NoStop}%
\bibitem [{\citenamefont {Boulebnane}\ and\ \citenamefont {Montanaro}(2024)}]{boulebnane2022solving}%
  \BibitemOpen
  \bibfield  {author} {\bibinfo {author} {\bibfnamefont {S.}~\bibnamefont {Boulebnane}}\ and\ \bibinfo {author} {\bibfnamefont {A.}~\bibnamefont {Montanaro}},\ }\href {https://doi.org/10.1103/PRXQuantum.5.030348} {\bibfield  {journal} {\bibinfo  {journal} {PRX Quantum}\ }\textbf {\bibinfo {volume} {5}},\ \bibinfo {pages} {030348} (\bibinfo {year} {2024})}\BibitemShut {NoStop}%
\bibitem [{\citenamefont {Kotil}\ \emph {et~al.}(2025)\citenamefont {Kotil}, \citenamefont {Pelofske}, \citenamefont {Riedmüller}, \citenamefont {Egger}, \citenamefont {Eidenbenz}, \citenamefont {Koch},\ and\ \citenamefont {Woerner}}]{kotil2025quantum}%
  \BibitemOpen
  \bibfield  {author} {\bibinfo {author} {\bibfnamefont {A.}~\bibnamefont {Kotil}}, \bibinfo {author} {\bibfnamefont {E.}~\bibnamefont {Pelofske}}, \bibinfo {author} {\bibfnamefont {S.}~\bibnamefont {Riedmüller}}, \bibinfo {author} {\bibfnamefont {D.~J.}\ \bibnamefont {Egger}}, \bibinfo {author} {\bibfnamefont {S.}~\bibnamefont {Eidenbenz}}, \bibinfo {author} {\bibfnamefont {T.}~\bibnamefont {Koch}},\ and\ \bibinfo {author} {\bibfnamefont {S.}~\bibnamefont {Woerner}},\ }\href {https://arxiv.org/abs/2503.22797} {\bibinfo {title} {{Quantum Approximate Multi-Objective Optimization}}} (\bibinfo {year} {2025}),\ \Eprint {https://arxiv.org/abs/2503.22797} {arXiv:2503.22797 [quant-ph]} \BibitemShut {NoStop}%
\bibitem [{\citenamefont {Koch}\ \emph {et~al.}(2025)\citenamefont {Koch}, \citenamefont {Neira}, \citenamefont {Chen}, \citenamefont {Cortiana}, \citenamefont {Egger}, \citenamefont {Heese}, \citenamefont {Hegade}, \citenamefont {Cadavid}, \citenamefont {Huang}, \citenamefont {Itoko} \emph {et~al.}}]{koch2025quantum}%
  \BibitemOpen
  \bibfield  {author} {\bibinfo {author} {\bibfnamefont {T.}~\bibnamefont {Koch}}, \bibinfo {author} {\bibfnamefont {D.~E.~B.}\ \bibnamefont {Neira}}, \bibinfo {author} {\bibfnamefont {Y.}~\bibnamefont {Chen}}, \bibinfo {author} {\bibfnamefont {G.}~\bibnamefont {Cortiana}}, \bibinfo {author} {\bibfnamefont {D.~J.}\ \bibnamefont {Egger}}, \bibinfo {author} {\bibfnamefont {R.}~\bibnamefont {Heese}}, \bibinfo {author} {\bibfnamefont {N.~N.}\ \bibnamefont {Hegade}}, \bibinfo {author} {\bibfnamefont {A.~G.}\ \bibnamefont {Cadavid}}, \bibinfo {author} {\bibfnamefont {R.}~\bibnamefont {Huang}}, \bibinfo {author} {\bibfnamefont {T.}~\bibnamefont {Itoko}}, \emph {et~al.},\ }\href {https://arxiv.org/abs/2504.03832} {\bibinfo {title} {{Quantum Optimization Benchmark Library -- The Intractable Decathlon}}} (\bibinfo {year} {2025}),\ \Eprint {https://arxiv.org/abs/2504.03832} {arXiv:2504.03832 [quant-ph]} \BibitemShut {NoStop}%
\bibitem [{\citenamefont {Durr}\ and\ \citenamefont {Hoyer}(1999)}]{durr1999quantumalgorithmfindingminimum}%
  \BibitemOpen
  \bibfield  {author} {\bibinfo {author} {\bibfnamefont {C.}~\bibnamefont {Durr}}\ and\ \bibinfo {author} {\bibfnamefont {P.}~\bibnamefont {Hoyer}},\ }\href {https://arxiv.org/abs/quant-ph/9607014} {\bibinfo {title} {{A Quantum Algorithm for Finding the Minimum}}} (\bibinfo {year} {1999}),\ \Eprint {https://arxiv.org/abs/quant-ph/9607014} {arXiv:quant-ph/9607014 [quant-ph]} \BibitemShut {NoStop}%
\bibitem [{\citenamefont {Somma}\ \emph {et~al.}(2008)\citenamefont {Somma}, \citenamefont {Boixo}, \citenamefont {Barnum},\ and\ \citenamefont {Knill}}]{somma2008quantum}%
  \BibitemOpen
  \bibfield  {author} {\bibinfo {author} {\bibfnamefont {R.~D.}\ \bibnamefont {Somma}}, \bibinfo {author} {\bibfnamefont {S.}~\bibnamefont {Boixo}}, \bibinfo {author} {\bibfnamefont {H.}~\bibnamefont {Barnum}},\ and\ \bibinfo {author} {\bibfnamefont {E.}~\bibnamefont {Knill}},\ }\href {https://doi.org/10.1103/PhysRevLett.101.130504} {\bibfield  {journal} {\bibinfo  {journal} {Phys. Rev. Lett.}\ }\textbf {\bibinfo {volume} {101}},\ \bibinfo {pages} {130504} (\bibinfo {year} {2008})}\BibitemShut {NoStop}%
\bibitem [{\citenamefont {Wocjan}\ and\ \citenamefont {Abeyesinghe}(2008)}]{wocjan2008speedup}%
  \BibitemOpen
  \bibfield  {author} {\bibinfo {author} {\bibfnamefont {P.}~\bibnamefont {Wocjan}}\ and\ \bibinfo {author} {\bibfnamefont {A.}~\bibnamefont {Abeyesinghe}},\ }\href {https://doi.org/10.1103/PhysRevA.78.042336} {\bibfield  {journal} {\bibinfo  {journal} {Phys. Rev. A}\ }\textbf {\bibinfo {volume} {78}},\ \bibinfo {pages} {042336} (\bibinfo {year} {2008})}\BibitemShut {NoStop}%
\bibitem [{\citenamefont {Hastings}(2018)}]{hastings2018shortpathquantum}%
  \BibitemOpen
  \bibfield  {author} {\bibinfo {author} {\bibfnamefont {M.~B.}\ \bibnamefont {Hastings}},\ }\href {https://doi.org/10.22331/q-2018-07-26-78} {\bibfield  {journal} {\bibinfo  {journal} {{Quantum}}\ }\textbf {\bibinfo {volume} {2}},\ \bibinfo {pages} {78} (\bibinfo {year} {2018})}\BibitemShut {NoStop}%
\bibitem [{\citenamefont {Montanaro}(2018)}]{montanaro2018quantum-walk}%
  \BibitemOpen
  \bibfield  {author} {\bibinfo {author} {\bibfnamefont {A.}~\bibnamefont {Montanaro}},\ }\href {https://doi.org/10.4086/toc.2018.v014a015} {\bibfield  {journal} {\bibinfo  {journal} {Theory of Computing}\ }\textbf {\bibinfo {volume} {14}},\ \bibinfo {pages} {1} (\bibinfo {year} {2018})}\BibitemShut {NoStop}%
\bibitem [{\citenamefont {Montanaro}(2020)}]{montanaro2020quantumspeedup}%
  \BibitemOpen
  \bibfield  {author} {\bibinfo {author} {\bibfnamefont {A.}~\bibnamefont {Montanaro}},\ }\href {https://doi.org/10.1103/PhysRevResearch.2.013056} {\bibfield  {journal} {\bibinfo  {journal} {Phys. Rev. Res.}\ }\textbf {\bibinfo {volume} {2}},\ \bibinfo {pages} {013056} (\bibinfo {year} {2020})}\BibitemShut {NoStop}%
\bibitem [{\citenamefont {Chakrabarti}\ \emph {et~al.}(2022)\citenamefont {Chakrabarti}, \citenamefont {Minssen}, \citenamefont {Yalovetzky},\ and\ \citenamefont {Pistoia}}]{chakrabarti2022universalquantumspeedupbranchandbound}%
  \BibitemOpen
  \bibfield  {author} {\bibinfo {author} {\bibfnamefont {S.}~\bibnamefont {Chakrabarti}}, \bibinfo {author} {\bibfnamefont {P.}~\bibnamefont {Minssen}}, \bibinfo {author} {\bibfnamefont {R.}~\bibnamefont {Yalovetzky}},\ and\ \bibinfo {author} {\bibfnamefont {M.}~\bibnamefont {Pistoia}},\ }\href {https://arxiv.org/abs/2210.03210} {\bibinfo {title} {{Universal Quantum Speedup for Branch-and-Bound, Branch-and-Cut, and Tree-Search Algorithms}}} (\bibinfo {year} {2022}),\ \Eprint {https://arxiv.org/abs/2210.03210} {arXiv:2210.03210 [quant-ph]} \BibitemShut {NoStop}%
\bibitem [{\citenamefont {Dalzell}\ \emph {et~al.}(2023)\citenamefont {Dalzell}, \citenamefont {Pancotti}, \citenamefont {Campbell},\ and\ \citenamefont {Brand\~{a}o}}]{dalzell2023mind}%
  \BibitemOpen
  \bibfield  {author} {\bibinfo {author} {\bibfnamefont {A.~M.}\ \bibnamefont {Dalzell}}, \bibinfo {author} {\bibfnamefont {N.}~\bibnamefont {Pancotti}}, \bibinfo {author} {\bibfnamefont {E.~T.}\ \bibnamefont {Campbell}},\ and\ \bibinfo {author} {\bibfnamefont {F.~G.}\ \bibnamefont {Brand\~{a}o}},\ }in\ \href {https://doi.org/10.1145/3564246.3585203} {\emph {\bibinfo {booktitle} {Proceedings of the 55th Annual ACM Symposium on Theory of Computing}}},\ \bibinfo {series and number} {STOC 2023}\ (\bibinfo  {publisher} {Association for Computing Machinery},\ \bibinfo {address} {New York, NY, USA},\ \bibinfo {year} {2023})\ p.\ \bibinfo {pages} {1131–1144}\BibitemShut {NoStop}%
\bibitem [{\citenamefont {Boehmer}(1967)}]{boehmer1967binary}%
  \BibitemOpen
  \bibfield  {author} {\bibinfo {author} {\bibfnamefont {A.}~\bibnamefont {Boehmer}},\ }\href {https://doi.org/10.1109/TIT.1967.1053969} {\bibfield  {journal} {\bibinfo  {journal} {IEEE Transactions on Information Theory}\ }\textbf {\bibinfo {volume} {13}},\ \bibinfo {pages} {156} (\bibinfo {year} {1967})}\BibitemShut {NoStop}%
\bibitem [{\citenamefont {Schroeder}(1970)}]{schroeder1970synthesis}%
  \BibitemOpen
  \bibfield  {author} {\bibinfo {author} {\bibfnamefont {M.}~\bibnamefont {Schroeder}},\ }\href {https://doi.org/10.1109/TIT.1970.1054411} {\bibfield  {journal} {\bibinfo  {journal} {IEEE Transactions on Information Theory}\ }\textbf {\bibinfo {volume} {16}},\ \bibinfo {pages} {85} (\bibinfo {year} {1970})}\BibitemShut {NoStop}%
\bibitem [{\citenamefont {Shaydulin}\ \emph {et~al.}(2024)\citenamefont {Shaydulin}, \citenamefont {Li}, \citenamefont {Chakrabarti}, \citenamefont {DeCross}, \citenamefont {Herman}, \citenamefont {Kumar}, \citenamefont {Larson}, \citenamefont {Lykov}, \citenamefont {Minssen}, \citenamefont {Sun} \emph {et~al.}}]{shaydulin2024evidence}%
  \BibitemOpen
  \bibfield  {author} {\bibinfo {author} {\bibfnamefont {R.}~\bibnamefont {Shaydulin}}, \bibinfo {author} {\bibfnamefont {C.}~\bibnamefont {Li}}, \bibinfo {author} {\bibfnamefont {S.}~\bibnamefont {Chakrabarti}}, \bibinfo {author} {\bibfnamefont {M.}~\bibnamefont {DeCross}}, \bibinfo {author} {\bibfnamefont {D.}~\bibnamefont {Herman}}, \bibinfo {author} {\bibfnamefont {N.}~\bibnamefont {Kumar}}, \bibinfo {author} {\bibfnamefont {J.}~\bibnamefont {Larson}}, \bibinfo {author} {\bibfnamefont {D.}~\bibnamefont {Lykov}}, \bibinfo {author} {\bibfnamefont {P.}~\bibnamefont {Minssen}}, \bibinfo {author} {\bibfnamefont {Y.}~\bibnamefont {Sun}}, \emph {et~al.},\ }\href {https://doi.org/10.1126/sciadv.adm6761} {\bibfield  {journal} {\bibinfo  {journal} {Science Advances}\ }\textbf {\bibinfo {volume} {10}},\ \bibinfo {pages} {eadm6761} (\bibinfo {year} {2024})}\BibitemShut {NoStop}%
\bibitem [{\citenamefont {Hegade}\ \emph {et~al.}(2021{\natexlab{a}})\citenamefont {Hegade}, \citenamefont {Paul}, \citenamefont {Albarr\'an-Arriagada}, \citenamefont {Chen},\ and\ \citenamefont {Solano}}]{hegade2021factorization}%
  \BibitemOpen
  \bibfield  {author} {\bibinfo {author} {\bibfnamefont {N.~N.}\ \bibnamefont {Hegade}}, \bibinfo {author} {\bibfnamefont {K.}~\bibnamefont {Paul}}, \bibinfo {author} {\bibfnamefont {F.}~\bibnamefont {Albarr\'an-Arriagada}}, \bibinfo {author} {\bibfnamefont {X.}~\bibnamefont {Chen}},\ and\ \bibinfo {author} {\bibfnamefont {E.}~\bibnamefont {Solano}},\ }\href {https://doi.org/10.1103/PhysRevA.104.L050403} {\bibfield  {journal} {\bibinfo  {journal} {Phys. Rev. A}\ }\textbf {\bibinfo {volume} {104}},\ \bibinfo {pages} {L050403} (\bibinfo {year} {2021}{\natexlab{a}})}\BibitemShut {NoStop}%
\bibitem [{\citenamefont {Battiti}(2009)}]{Battiti2009}%
  \BibitemOpen
  \bibfield  {author} {\bibinfo {author} {\bibfnamefont {R.}~\bibnamefont {Battiti}},\ }\bibinfo {title} {Maximum satisfiability problem},\ in\ \href {https://doi.org/10.1007/978-0-387-74759-0_364} {\emph {\bibinfo {booktitle} {Encyclopedia of Optimization}}},\ \bibinfo {editor} {edited by\ \bibinfo {editor} {\bibfnamefont {C.~A.}\ \bibnamefont {Floudas}}\ and\ \bibinfo {editor} {\bibfnamefont {P.~M.}\ \bibnamefont {Pardalos}}}\ (\bibinfo  {publisher} {Springer US},\ \bibinfo {address} {Boston, MA},\ \bibinfo {year} {2009})\ pp.\ \bibinfo {pages} {2035--2041}\BibitemShut {NoStop}%
\bibitem [{\citenamefont {Barron}\ \emph {et~al.}(2024)\citenamefont {Barron}, \citenamefont {Egger}, \citenamefont {Pelofske}, \citenamefont {Bärtschi}, \citenamefont {Eidenbenz}, \citenamefont {Lehmkuehler},\ and\ \citenamefont {Woerner}}]{barron2023provableboundsnoisefreeexpectation}%
  \BibitemOpen
  \bibfield  {author} {\bibinfo {author} {\bibfnamefont {S.~V.}\ \bibnamefont {Barron}}, \bibinfo {author} {\bibfnamefont {D.~J.}\ \bibnamefont {Egger}}, \bibinfo {author} {\bibfnamefont {E.}~\bibnamefont {Pelofske}}, \bibinfo {author} {\bibfnamefont {A.}~\bibnamefont {Bärtschi}}, \bibinfo {author} {\bibfnamefont {S.}~\bibnamefont {Eidenbenz}}, \bibinfo {author} {\bibfnamefont {M.}~\bibnamefont {Lehmkuehler}},\ and\ \bibinfo {author} {\bibfnamefont {S.}~\bibnamefont {Woerner}},\ }\href {https://doi.org/10.1038/s43588-024-00709-1} {\bibfield  {journal} {\bibinfo  {journal} {Nature Computational Science}\ }\textbf {\bibinfo {volume} {4}},\ \bibinfo {pages} {865–875} (\bibinfo {year} {2024})}\BibitemShut {NoStop}%
\bibitem [{\citenamefont {Perdomo}\ \emph {et~al.}(2008)\citenamefont {Perdomo}, \citenamefont {Truncik}, \citenamefont {Tubert-Brohman}, \citenamefont {Rose},\ and\ \citenamefont {Aspuru-Guzik}}]{perdomo2008construction}%
  \BibitemOpen
  \bibfield  {author} {\bibinfo {author} {\bibfnamefont {A.}~\bibnamefont {Perdomo}}, \bibinfo {author} {\bibfnamefont {C.}~\bibnamefont {Truncik}}, \bibinfo {author} {\bibfnamefont {I.}~\bibnamefont {Tubert-Brohman}}, \bibinfo {author} {\bibfnamefont {G.}~\bibnamefont {Rose}},\ and\ \bibinfo {author} {\bibfnamefont {A.}~\bibnamefont {Aspuru-Guzik}},\ }\href {https://doi.org/10.1103/PhysRevA.78.012320} {\bibfield  {journal} {\bibinfo  {journal} {Phys. Rev. A}\ }\textbf {\bibinfo {volume} {78}},\ \bibinfo {pages} {012320} (\bibinfo {year} {2008})}\BibitemShut {NoStop}%
\bibitem [{\citenamefont {Perdomo-Ortiz}\ \emph {et~al.}(2012)\citenamefont {Perdomo-Ortiz}, \citenamefont {Dickson}, \citenamefont {Drew-Brook}, \citenamefont {Rose},\ and\ \citenamefont {Aspuru-Guzik}}]{perdomo2012finding}%
  \BibitemOpen
  \bibfield  {author} {\bibinfo {author} {\bibfnamefont {A.}~\bibnamefont {Perdomo-Ortiz}}, \bibinfo {author} {\bibfnamefont {N.}~\bibnamefont {Dickson}}, \bibinfo {author} {\bibfnamefont {M.}~\bibnamefont {Drew-Brook}}, \bibinfo {author} {\bibfnamefont {G.}~\bibnamefont {Rose}},\ and\ \bibinfo {author} {\bibfnamefont {A.}~\bibnamefont {Aspuru-Guzik}},\ }\href {https://doi.org/10.1038/srep00571} {\bibfield  {journal} {\bibinfo  {journal} {Scientific Reports}\ }\textbf {\bibinfo {volume} {2}},\ \bibinfo {pages} {571} (\bibinfo {year} {2012})}\BibitemShut {NoStop}%
\bibitem [{\citenamefont {Babbush}\ \emph {et~al.}(2014)\citenamefont {Babbush}, \citenamefont {Perdomo-Ortiz}, \citenamefont {O'Gorman}, \citenamefont {Macready},\ and\ \citenamefont {Aspuru-Guzik}}]{babbush2014construction}%
  \BibitemOpen
  \bibfield  {author} {\bibinfo {author} {\bibfnamefont {R.}~\bibnamefont {Babbush}}, \bibinfo {author} {\bibfnamefont {A.}~\bibnamefont {Perdomo-Ortiz}}, \bibinfo {author} {\bibfnamefont {B.}~\bibnamefont {O'Gorman}}, \bibinfo {author} {\bibfnamefont {W.}~\bibnamefont {Macready}},\ and\ \bibinfo {author} {\bibfnamefont {A.}~\bibnamefont {Aspuru-Guzik}},\ }\bibinfo {title} {{Construction of Energy Functions for Lattice Heteropolymer Models: Efficient Encodings for Constraint Satisfaction Programming and Quantum Annealing}},\ in\ \href {https://doi.org/https://doi.org/10.1002/9781118755815.ch05} {\emph {\bibinfo {booktitle} {{Advances in Chemical Physics: Volume 155}}}}\ (\bibinfo  {publisher} {John Wiley \& Sons, Ltd},\ \bibinfo {year} {2014})\ Chap.~\bibinfo {chapter} {5}, pp.\ \bibinfo {pages} {201--244}\BibitemShut {NoStop}%
\bibitem [{\citenamefont {Babej}\ \emph {et~al.}(2018)\citenamefont {Babej}, \citenamefont {Ing},\ and\ \citenamefont {Fingerhuth}}]{babej2018coarsegrainedlatticeproteinfolding}%
  \BibitemOpen
  \bibfield  {author} {\bibinfo {author} {\bibfnamefont {T.}~\bibnamefont {Babej}}, \bibinfo {author} {\bibfnamefont {C.}~\bibnamefont {Ing}},\ and\ \bibinfo {author} {\bibfnamefont {M.}~\bibnamefont {Fingerhuth}},\ }\href {https://arxiv.org/abs/1811.00713} {\bibinfo {title} {Coarse-grained lattice protein folding on a quantum annealer}} (\bibinfo {year} {2018}),\ \Eprint {https://arxiv.org/abs/1811.00713} {arXiv:1811.00713 [quant-ph]} \BibitemShut {NoStop}%
\bibitem [{\citenamefont {Robert}\ \emph {et~al.}(2021)\citenamefont {Robert}, \citenamefont {Barkoutsos}, \citenamefont {Woerner},\ and\ \citenamefont {Tavernelli}}]{robert2021resource}%
  \BibitemOpen
  \bibfield  {author} {\bibinfo {author} {\bibfnamefont {A.}~\bibnamefont {Robert}}, \bibinfo {author} {\bibfnamefont {P.~K.}\ \bibnamefont {Barkoutsos}}, \bibinfo {author} {\bibfnamefont {S.}~\bibnamefont {Woerner}},\ and\ \bibinfo {author} {\bibfnamefont {I.}~\bibnamefont {Tavernelli}},\ }\href {https://doi.org/10.1038/s41534-021-00368-4} {\bibfield  {journal} {\bibinfo  {journal} {npj Quantum Information}\ }\textbf {\bibinfo {volume} {7}},\ \bibinfo {pages} {1–5} (\bibinfo {year} {2021})}\BibitemShut {NoStop}%
\bibitem [{\citenamefont {Chandarana}\ \emph {et~al.}(2023)\citenamefont {Chandarana}, \citenamefont {Hegade}, \citenamefont {Montalban}, \citenamefont {Solano},\ and\ \citenamefont {Chen}}]{chandarana2023digitized}%
  \BibitemOpen
  \bibfield  {author} {\bibinfo {author} {\bibfnamefont {P.}~\bibnamefont {Chandarana}}, \bibinfo {author} {\bibfnamefont {N.~N.}\ \bibnamefont {Hegade}}, \bibinfo {author} {\bibfnamefont {I.}~\bibnamefont {Montalban}}, \bibinfo {author} {\bibfnamefont {E.}~\bibnamefont {Solano}},\ and\ \bibinfo {author} {\bibfnamefont {X.}~\bibnamefont {Chen}},\ }\href {https://doi.org/10.1103/PhysRevApplied.20.014024} {\bibfield  {journal} {\bibinfo  {journal} {Phys. Rev. Appl.}\ }\textbf {\bibinfo {volume} {20}},\ \bibinfo {pages} {014024} (\bibinfo {year} {2023})}\BibitemShut {NoStop}%
\bibitem [{\citenamefont {Pamidimukkala}\ \emph {et~al.}(2024)\citenamefont {Pamidimukkala}, \citenamefont {Bopardikar}, \citenamefont {Dakshinamoorthy}, \citenamefont {Kannan}, \citenamefont {Dasgupta},\ and\ \citenamefont {Senapati}}]{pamidimukkala2024protein}%
  \BibitemOpen
  \bibfield  {author} {\bibinfo {author} {\bibfnamefont {J.~V.}\ \bibnamefont {Pamidimukkala}}, \bibinfo {author} {\bibfnamefont {S.}~\bibnamefont {Bopardikar}}, \bibinfo {author} {\bibfnamefont {A.}~\bibnamefont {Dakshinamoorthy}}, \bibinfo {author} {\bibfnamefont {A.}~\bibnamefont {Kannan}}, \bibinfo {author} {\bibfnamefont {K.}~\bibnamefont {Dasgupta}},\ and\ \bibinfo {author} {\bibfnamefont {S.}~\bibnamefont {Senapati}},\ }\href {https://doi.org/10.1021/acs.jctc.4c00848} {\bibfield  {journal} {\bibinfo  {journal} {Journal of Chemical Theory and Computation}\ }\textbf {\bibinfo {volume} {20}},\ \bibinfo {pages} {10223} (\bibinfo {year} {2024})}\BibitemShut {NoStop}%
\bibitem [{\citenamefont {Wang}\ and\ \citenamefont {Zhou}(2024)}]{wang2025efficient}%
  \BibitemOpen
  \bibfield  {author} {\bibinfo {author} {\bibfnamefont {Y.}~\bibnamefont {Wang}}\ and\ \bibinfo {author} {\bibfnamefont {X.}~\bibnamefont {Zhou}},\ }\href {https://doi.org/10.1088/2058-9565/ada08e} {\bibfield  {journal} {\bibinfo  {journal} {Quantum Science and Technology}\ }\textbf {\bibinfo {volume} {10}},\ \bibinfo {pages} {015056} (\bibinfo {year} {2024})}\BibitemShut {NoStop}%
\bibitem [{\citenamefont {Demirplak}\ and\ \citenamefont {Rice}(2003)}]{demirplak2003adiabatic}%
  \BibitemOpen
  \bibfield  {author} {\bibinfo {author} {\bibfnamefont {M.}~\bibnamefont {Demirplak}}\ and\ \bibinfo {author} {\bibfnamefont {S.~A.}\ \bibnamefont {Rice}},\ }\href {https://pubs.acs.org/doi/10.1021/jp030708a} {\bibfield  {journal} {\bibinfo  {journal} {The Journal of Physical Chemistry A}\ }\textbf {\bibinfo {volume} {107}},\ \bibinfo {pages} {9937} (\bibinfo {year} {2003})}\BibitemShut {NoStop}%
\bibitem [{\citenamefont {Berry}(2009)}]{berry2009transitionless}%
  \BibitemOpen
  \bibfield  {author} {\bibinfo {author} {\bibfnamefont {M.~V.}\ \bibnamefont {Berry}},\ }\href {https://iopscience.iop.org/article/10.1088/1751-8113/42/36/365303} {\bibfield  {journal} {\bibinfo  {journal} {Journal of Physics A: Mathematical and Theoretical}\ }\textbf {\bibinfo {volume} {42}},\ \bibinfo {pages} {365303} (\bibinfo {year} {2009})}\BibitemShut {NoStop}%
\bibitem [{\citenamefont {Chen}\ \emph {et~al.}(2010)\citenamefont {Chen}, \citenamefont {Ruschhaupt}, \citenamefont {Schmidt}, \citenamefont {del Campo}, \citenamefont {Gu\'ery-Odelin},\ and\ \citenamefont {Muga}}]{chen2010fast}%
  \BibitemOpen
  \bibfield  {author} {\bibinfo {author} {\bibfnamefont {X.}~\bibnamefont {Chen}}, \bibinfo {author} {\bibfnamefont {A.}~\bibnamefont {Ruschhaupt}}, \bibinfo {author} {\bibfnamefont {S.}~\bibnamefont {Schmidt}}, \bibinfo {author} {\bibfnamefont {A.}~\bibnamefont {del Campo}}, \bibinfo {author} {\bibfnamefont {D.}~\bibnamefont {Gu\'ery-Odelin}},\ and\ \bibinfo {author} {\bibfnamefont {J.~G.}\ \bibnamefont {Muga}},\ }\href {https://doi.org/10.1103/PhysRevLett.104.063002} {\bibfield  {journal} {\bibinfo  {journal} {Phys. Rev. Lett.}\ }\textbf {\bibinfo {volume} {104}},\ \bibinfo {pages} {063002} (\bibinfo {year} {2010})}\BibitemShut {NoStop}%
\bibitem [{\citenamefont {del Campo}(2013)}]{campo2013shortcuts}%
  \BibitemOpen
  \bibfield  {author} {\bibinfo {author} {\bibfnamefont {A.}~\bibnamefont {del Campo}},\ }\href {https://doi.org/10.1103/PhysRevLett.111.100502} {\bibfield  {journal} {\bibinfo  {journal} {Phys. Rev. Lett.}\ }\textbf {\bibinfo {volume} {111}},\ \bibinfo {pages} {100502} (\bibinfo {year} {2013})}\BibitemShut {NoStop}%
\bibitem [{\citenamefont {Sels}\ and\ \citenamefont {Polkovnikov}(2017)}]{sels2017minimizing}%
  \BibitemOpen
  \bibfield  {author} {\bibinfo {author} {\bibfnamefont {D.}~\bibnamefont {Sels}}\ and\ \bibinfo {author} {\bibfnamefont {A.}~\bibnamefont {Polkovnikov}},\ }\href {https://www.pnas.org/doi/full/10.1073/pnas.1619826114} {\bibfield  {journal} {\bibinfo  {journal} {Proceedings of the National Academy of Sciences}\ }\textbf {\bibinfo {volume} {114}},\ \bibinfo {pages} {E3909} (\bibinfo {year} {2017})}\BibitemShut {NoStop}%
\bibitem [{\citenamefont {Claeys}\ \emph {et~al.}(2019)\citenamefont {Claeys}, \citenamefont {Pandey}, \citenamefont {Sels},\ and\ \citenamefont {Polkovnikov}}]{claeys2019floquet}%
  \BibitemOpen
  \bibfield  {author} {\bibinfo {author} {\bibfnamefont {P.~W.}\ \bibnamefont {Claeys}}, \bibinfo {author} {\bibfnamefont {M.}~\bibnamefont {Pandey}}, \bibinfo {author} {\bibfnamefont {D.}~\bibnamefont {Sels}},\ and\ \bibinfo {author} {\bibfnamefont {A.}~\bibnamefont {Polkovnikov}},\ }\href {https://journals.aps.org/prl/abstract/10.1103/PhysRevLett.123.090602} {\bibfield  {journal} {\bibinfo  {journal} {Phys. Rev. Lett.}\ }\textbf {\bibinfo {volume} {123}},\ \bibinfo {pages} {090602} (\bibinfo {year} {2019})}\BibitemShut {NoStop}%
\bibitem [{\citenamefont {Takahashi}\ and\ \citenamefont {del Campo}(2024)}]{takahashi2024shortcuts}%
  \BibitemOpen
  \bibfield  {author} {\bibinfo {author} {\bibfnamefont {K.}~\bibnamefont {Takahashi}}\ and\ \bibinfo {author} {\bibfnamefont {A.}~\bibnamefont {del Campo}},\ }\href {https://journals.aps.org/prx/abstract/10.1103/PhysRevX.14.011032} {\bibfield  {journal} {\bibinfo  {journal} {Phys. Rev. X}\ }\textbf {\bibinfo {volume} {14}},\ \bibinfo {pages} {011032} (\bibinfo {year} {2024})}\BibitemShut {NoStop}%
\bibitem [{\citenamefont {Chandarana}\ \emph {et~al.}(2022)\citenamefont {Chandarana}, \citenamefont {Hegade}, \citenamefont {Paul}, \citenamefont {Albarr\'an-Arriagada}, \citenamefont {Solano}, \citenamefont {del Campo},\ and\ \citenamefont {Chen}}]{chandarana2022digitized}%
  \BibitemOpen
  \bibfield  {author} {\bibinfo {author} {\bibfnamefont {P.}~\bibnamefont {Chandarana}}, \bibinfo {author} {\bibfnamefont {N.~N.}\ \bibnamefont {Hegade}}, \bibinfo {author} {\bibfnamefont {K.}~\bibnamefont {Paul}}, \bibinfo {author} {\bibfnamefont {F.}~\bibnamefont {Albarr\'an-Arriagada}}, \bibinfo {author} {\bibfnamefont {E.}~\bibnamefont {Solano}}, \bibinfo {author} {\bibfnamefont {A.}~\bibnamefont {del Campo}},\ and\ \bibinfo {author} {\bibfnamefont {X.}~\bibnamefont {Chen}},\ }\href {https://doi.org/10.1103/PhysRevResearch.4.013141} {\bibfield  {journal} {\bibinfo  {journal} {Phys. Rev. Res.}\ }\textbf {\bibinfo {volume} {4}},\ \bibinfo {pages} {013141} (\bibinfo {year} {2022})}\BibitemShut {NoStop}%
\bibitem [{\citenamefont {Hegade}\ \emph {et~al.}(2022)\citenamefont {Hegade}, \citenamefont {Chen},\ and\ \citenamefont {Solano}}]{hegade2022digitized}%
  \BibitemOpen
  \bibfield  {author} {\bibinfo {author} {\bibfnamefont {N.~N.}\ \bibnamefont {Hegade}}, \bibinfo {author} {\bibfnamefont {X.}~\bibnamefont {Chen}},\ and\ \bibinfo {author} {\bibfnamefont {E.}~\bibnamefont {Solano}},\ }\href {https://journals.aps.org/prresearch/abstract/10.1103/PhysRevResearch.4.L042030} {\bibfield  {journal} {\bibinfo  {journal} {Phys. Rev. Res.}\ }\textbf {\bibinfo {volume} {4}},\ \bibinfo {pages} {L042030} (\bibinfo {year} {2022})}\BibitemShut {NoStop}%
\bibitem [{\citenamefont {Simen}\ \emph {et~al.}(2025)\citenamefont {Simen}, \citenamefont {Romero}, \citenamefont {Cadavid}, \citenamefont {Solano},\ and\ \citenamefont {Hegade}}]{simen2025branch}%
  \BibitemOpen
  \bibfield  {author} {\bibinfo {author} {\bibfnamefont {A.}~\bibnamefont {Simen}}, \bibinfo {author} {\bibfnamefont {S.~V.}\ \bibnamefont {Romero}}, \bibinfo {author} {\bibfnamefont {A.~G.}\ \bibnamefont {Cadavid}}, \bibinfo {author} {\bibfnamefont {E.}~\bibnamefont {Solano}},\ and\ \bibinfo {author} {\bibfnamefont {N.~N.}\ \bibnamefont {Hegade}},\ }\href {https://arxiv.org/abs/2504.15367} {\bibinfo {title} {Branch-and-bound digitized counterdiabatic quantum optimization}} (\bibinfo {year} {2025}),\ \Eprint {https://arxiv.org/abs/2504.15367} {arXiv:2504.15367 [quant-ph]} \BibitemShut {NoStop}%
\bibitem [{\citenamefont {Cadavid}\ \emph {et~al.}(2025)\citenamefont {Cadavid}, \citenamefont {Dalal}, \citenamefont {Simen}, \citenamefont {Solano},\ and\ \citenamefont {Hegade}}]{cadavid2024bias}%
  \BibitemOpen
  \bibfield  {author} {\bibinfo {author} {\bibfnamefont {A.~G.}\ \bibnamefont {Cadavid}}, \bibinfo {author} {\bibfnamefont {A.}~\bibnamefont {Dalal}}, \bibinfo {author} {\bibfnamefont {A.}~\bibnamefont {Simen}}, \bibinfo {author} {\bibfnamefont {E.}~\bibnamefont {Solano}},\ and\ \bibinfo {author} {\bibfnamefont {N.~N.}\ \bibnamefont {Hegade}},\ }\href {https://doi.org/10.1103/PhysRevResearch.7.L022010} {\bibfield  {journal} {\bibinfo  {journal} {Phys. Rev. Res.}\ }\textbf {\bibinfo {volume} {7}},\ \bibinfo {pages} {L022010} (\bibinfo {year} {2025})}\BibitemShut {NoStop}%
\bibitem [{\citenamefont {Romero}\ \emph {et~al.}(2024)\citenamefont {Romero}, \citenamefont {Visuri}, \citenamefont {Cadavid}, \citenamefont {Solano},\ and\ \citenamefont {Hegade}}]{romero2024bias}%
  \BibitemOpen
  \bibfield  {author} {\bibinfo {author} {\bibfnamefont {S.~V.}\ \bibnamefont {Romero}}, \bibinfo {author} {\bibfnamefont {A.-M.}\ \bibnamefont {Visuri}}, \bibinfo {author} {\bibfnamefont {A.~G.}\ \bibnamefont {Cadavid}}, \bibinfo {author} {\bibfnamefont {E.}~\bibnamefont {Solano}},\ and\ \bibinfo {author} {\bibfnamefont {N.~N.}\ \bibnamefont {Hegade}},\ }\href {https://arxiv.org/abs/2409.04477} {\bibinfo {title} {{Bias-Field Digitized Counterdiabatic Quantum Algorithm for Higher-Order Binary Optimization}}} (\bibinfo {year} {2024}),\ \Eprint {https://arxiv.org/abs/2409.04477} {arXiv:2409.04477 [quant-ph]} \BibitemShut {NoStop}%
\bibitem [{\citenamefont {{IBM Quantum}}(2025)}]{iskay}%
  \BibitemOpen
  \bibfield  {author} {\bibinfo {author} {\bibnamefont {{IBM Quantum}}},\ }\href@noop {} {\bibinfo {title} {{Iskay Quantum Optimizer - A Qiskit Function by Kipu Quantum}}},\ \bibinfo {howpublished} {\url{https://docs.quantum.ibm.com/guides/kipu-optimization}} (\bibinfo {year} {2025}),\ \bibinfo {note} {[Online: 14/04/25]}\BibitemShut {NoStop}%
\bibitem [{\citenamefont {Gra{\ss}}(2019)}]{grass2019quantum}%
  \BibitemOpen
  \bibfield  {author} {\bibinfo {author} {\bibfnamefont {T.}~\bibnamefont {Gra{\ss}}},\ }\href {https://journals.aps.org/prl/abstract/10.1103/PhysRevLett.123.120501} {\bibfield  {journal} {\bibinfo  {journal} {Phys. Rev. Lett.}\ }\textbf {\bibinfo {volume} {123}},\ \bibinfo {pages} {120501} (\bibinfo {year} {2019})}\BibitemShut {NoStop}%
\bibitem [{\citenamefont {Grass}(2022)}]{grass2022quantum}%
  \BibitemOpen
  \bibfield  {author} {\bibinfo {author} {\bibfnamefont {T.}~\bibnamefont {Grass}},\ }\href {https://journals.aps.org/prapplied/abstract/10.1103/PhysRevApplied.18.044036} {\bibfield  {journal} {\bibinfo  {journal} {Phys. Rev. Appl.}\ }\textbf {\bibinfo {volume} {18}},\ \bibinfo {pages} {044036} (\bibinfo {year} {2022})}\BibitemShut {NoStop}%
\bibitem [{\citenamefont {Cerezo}\ \emph {et~al.}(2024)\citenamefont {Cerezo}, \citenamefont {Larocca}, \citenamefont {García-Martín}, \citenamefont {Diaz}, \citenamefont {Braccia}, \citenamefont {Fontana}, \citenamefont {Rudolph}, \citenamefont {Bermejo}, \citenamefont {Ijaz}, \citenamefont {Thanasilp} \emph {et~al.}}]{cerezo2024doesprovableabsencebarren}%
  \BibitemOpen
  \bibfield  {author} {\bibinfo {author} {\bibfnamefont {M.}~\bibnamefont {Cerezo}}, \bibinfo {author} {\bibfnamefont {M.}~\bibnamefont {Larocca}}, \bibinfo {author} {\bibfnamefont {D.}~\bibnamefont {García-Martín}}, \bibinfo {author} {\bibfnamefont {N.~L.}\ \bibnamefont {Diaz}}, \bibinfo {author} {\bibfnamefont {P.}~\bibnamefont {Braccia}}, \bibinfo {author} {\bibfnamefont {E.}~\bibnamefont {Fontana}}, \bibinfo {author} {\bibfnamefont {M.~S.}\ \bibnamefont {Rudolph}}, \bibinfo {author} {\bibfnamefont {P.}~\bibnamefont {Bermejo}}, \bibinfo {author} {\bibfnamefont {A.}~\bibnamefont {Ijaz}}, \bibinfo {author} {\bibfnamefont {S.}~\bibnamefont {Thanasilp}}, \emph {et~al.},\ }\href {https://arxiv.org/abs/2312.09121} {\bibinfo {title} {{Does provable absence of barren plateaus imply classical simulability? Or, why we need to rethink variational quantum computing}}} (\bibinfo {year} {2024}),\ \Eprint {https://arxiv.org/abs/2312.09121} {arXiv:2312.09121 [quant-ph]} \BibitemShut {NoStop}%
\bibitem [{\citenamefont {Larocca}\ \emph {et~al.}(2025)\citenamefont {Larocca}, \citenamefont {Thanasilp}, \citenamefont {Wang}, \citenamefont {Sharma}, \citenamefont {Biamonte}, \citenamefont {Coles}, \citenamefont {Cincio}, \citenamefont {McClean}, \citenamefont {Holmes},\ and\ \citenamefont {Cerezo}}]{larocca2024reviewbarrenplateausvariational}%
  \BibitemOpen
  \bibfield  {author} {\bibinfo {author} {\bibfnamefont {M.}~\bibnamefont {Larocca}}, \bibinfo {author} {\bibfnamefont {S.}~\bibnamefont {Thanasilp}}, \bibinfo {author} {\bibfnamefont {S.}~\bibnamefont {Wang}}, \bibinfo {author} {\bibfnamefont {K.}~\bibnamefont {Sharma}}, \bibinfo {author} {\bibfnamefont {J.}~\bibnamefont {Biamonte}}, \bibinfo {author} {\bibfnamefont {P.~J.}\ \bibnamefont {Coles}}, \bibinfo {author} {\bibfnamefont {L.}~\bibnamefont {Cincio}}, \bibinfo {author} {\bibfnamefont {J.~R.}\ \bibnamefont {McClean}}, \bibinfo {author} {\bibfnamefont {Z.}~\bibnamefont {Holmes}},\ and\ \bibinfo {author} {\bibfnamefont {M.}~\bibnamefont {Cerezo}},\ }\href {https://doi.org/10.1038/s42254-025-00813-9} {\bibfield  {journal} {\bibinfo  {journal} {Nature Reviews Physics}\ }\textbf {\bibinfo {volume} {7}},\ \bibinfo {pages} {174–189} (\bibinfo {year} {2025})}\BibitemShut {NoStop}%
\bibitem [{\citenamefont {Chandarana}\ \emph {et~al.}(2025)\citenamefont {Chandarana}, \citenamefont {Cadavid}, \citenamefont {Romero}, \citenamefont {Simen}, \citenamefont {Solano},\ and\ \citenamefont {Hegade}}]{chandarana2025runtime}%
  \BibitemOpen
  \bibfield  {author} {\bibinfo {author} {\bibfnamefont {P.}~\bibnamefont {Chandarana}}, \bibinfo {author} {\bibfnamefont {A.~G.}\ \bibnamefont {Cadavid}}, \bibinfo {author} {\bibfnamefont {S.~V.}\ \bibnamefont {Romero}}, \bibinfo {author} {\bibfnamefont {A.}~\bibnamefont {Simen}}, \bibinfo {author} {\bibfnamefont {E.}~\bibnamefont {Solano}},\ and\ \bibinfo {author} {\bibfnamefont {N.~N.}\ \bibnamefont {Hegade}},\ }\href {https://arxiv.org/abs/2505.08663} {\bibinfo {title} {{Runtime Quantum Advantage with Digital Quantum Optimization}}} (\bibinfo {year} {2025}),\ \Eprint {https://arxiv.org/abs/2505.08663} {arXiv:2505.08663 [quant-ph]} \BibitemShut {NoStop}%
\bibitem [{\citenamefont {{IonQ}}(2025)}]{ionq}%
  \BibitemOpen
  \bibfield  {author} {\bibinfo {author} {\bibnamefont {{IonQ}}},\ }\href {https://ionq.com/} {}\bibinfo {howpublished} {\url{https://ionq.com/}} (\bibinfo {year} {2025})\BibitemShut {NoStop}%
\bibitem [{\citenamefont {Sherrington}\ and\ \citenamefont {Kirkpatrick}(1975)}]{sherrington1975solvable}%
  \BibitemOpen
  \bibfield  {author} {\bibinfo {author} {\bibfnamefont {D.}~\bibnamefont {Sherrington}}\ and\ \bibinfo {author} {\bibfnamefont {S.}~\bibnamefont {Kirkpatrick}},\ }\href {https://doi.org/10.1103/PhysRevLett.35.1792} {\bibfield  {journal} {\bibinfo  {journal} {Phys. Rev. Lett.}\ }\textbf {\bibinfo {volume} {35}},\ \bibinfo {pages} {1792} (\bibinfo {year} {1975})}\BibitemShut {NoStop}%
\bibitem [{\citenamefont {{D-Wave Systems}}(2025)}]{dwave}%
  \BibitemOpen
  \bibfield  {author} {\bibinfo {author} {\bibnamefont {{D-Wave Systems}}},\ }\href {https://www.dwavesys.com/} {}\bibinfo {howpublished} {\url{https://www.dwavesys.com/}} (\bibinfo {year} {2025})\BibitemShut {NoStop}%
\bibitem [{\citenamefont {Pelofske}\ \emph {et~al.}(2023{\natexlab{b}})\citenamefont {Pelofske}, \citenamefont {Bärtschi}, \citenamefont {Golden},\ and\ \citenamefont {Eidenbenz}}]{pelofske2023highround}%
  \BibitemOpen
  \bibfield  {author} {\bibinfo {author} {\bibfnamefont {E.}~\bibnamefont {Pelofske}}, \bibinfo {author} {\bibfnamefont {A.}~\bibnamefont {Bärtschi}}, \bibinfo {author} {\bibfnamefont {J.}~\bibnamefont {Golden}},\ and\ \bibinfo {author} {\bibfnamefont {S.}~\bibnamefont {Eidenbenz}},\ }in\ \href {https://doi.org/10.1109/QCE57702.2023.00064} {\emph {\bibinfo {booktitle} {2023 IEEE International Conference on Quantum Computing and Engineering (QCE)}}},\ Vol.~\bibinfo {volume} {01}\ (\bibinfo {year} {2023})\ pp.\ \bibinfo {pages} {506--517}\BibitemShut {NoStop}%
\bibitem [{\citenamefont {Kolodrubetz}\ \emph {et~al.}(2017)\citenamefont {Kolodrubetz}, \citenamefont {Sels}, \citenamefont {Mehta},\ and\ \citenamefont {Polkovnikov}}]{kolodrubetz2017geometry}%
  \BibitemOpen
  \bibfield  {author} {\bibinfo {author} {\bibfnamefont {M.}~\bibnamefont {Kolodrubetz}}, \bibinfo {author} {\bibfnamefont {D.}~\bibnamefont {Sels}}, \bibinfo {author} {\bibfnamefont {P.}~\bibnamefont {Mehta}},\ and\ \bibinfo {author} {\bibfnamefont {A.}~\bibnamefont {Polkovnikov}},\ }\href {https://www.sciencedirect.com/science/article/abs/pii/S0370157317301989} {\bibfield  {journal} {\bibinfo  {journal} {Physics Reports}\ }\textbf {\bibinfo {volume} {697}},\ \bibinfo {pages} {1} (\bibinfo {year} {2017})}\BibitemShut {NoStop}%
\bibitem [{\citenamefont {Hatomura}\ and\ \citenamefont {Takahashi}(2021)}]{hatomura2021controlling}%
  \BibitemOpen
  \bibfield  {author} {\bibinfo {author} {\bibfnamefont {T.}~\bibnamefont {Hatomura}}\ and\ \bibinfo {author} {\bibfnamefont {K.}~\bibnamefont {Takahashi}},\ }\href {https://journals.aps.org/pra/abstract/10.1103/PhysRevA.103.012220} {\bibfield  {journal} {\bibinfo  {journal} {Phys. Rev. A}\ }\textbf {\bibinfo {volume} {103}},\ \bibinfo {pages} {012220} (\bibinfo {year} {2021})}\BibitemShut {NoStop}%
\bibitem [{\citenamefont {Hormozi}\ \emph {et~al.}(2017)\citenamefont {Hormozi}, \citenamefont {Brown}, \citenamefont {Carleo},\ and\ \citenamefont {Troyer}}]{hormozi2017nonstoquastic}%
  \BibitemOpen
  \bibfield  {author} {\bibinfo {author} {\bibfnamefont {L.}~\bibnamefont {Hormozi}}, \bibinfo {author} {\bibfnamefont {E.~W.}\ \bibnamefont {Brown}}, \bibinfo {author} {\bibfnamefont {G.}~\bibnamefont {Carleo}},\ and\ \bibinfo {author} {\bibfnamefont {M.}~\bibnamefont {Troyer}},\ }\href {https://doi.org/10.1103/PhysRevB.95.184416} {\bibfield  {journal} {\bibinfo  {journal} {Phys. Rev. B}\ }\textbf {\bibinfo {volume} {95}},\ \bibinfo {pages} {184416} (\bibinfo {year} {2017})}\BibitemShut {NoStop}%
\bibitem [{\citenamefont {Hegade}\ \emph {et~al.}(2021{\natexlab{b}})\citenamefont {Hegade}, \citenamefont {Paul}, \citenamefont {Ding}, \citenamefont {Sanz}, \citenamefont {Albarr{\'a}n-Arriagada}, \citenamefont {Solano},\ and\ \citenamefont {Chen}}]{hegade2021shortcuts}%
  \BibitemOpen
  \bibfield  {author} {\bibinfo {author} {\bibfnamefont {N.~N.}\ \bibnamefont {Hegade}}, \bibinfo {author} {\bibfnamefont {K.}~\bibnamefont {Paul}}, \bibinfo {author} {\bibfnamefont {Y.}~\bibnamefont {Ding}}, \bibinfo {author} {\bibfnamefont {M.}~\bibnamefont {Sanz}}, \bibinfo {author} {\bibfnamefont {F.}~\bibnamefont {Albarr{\'a}n-Arriagada}}, \bibinfo {author} {\bibfnamefont {E.}~\bibnamefont {Solano}},\ and\ \bibinfo {author} {\bibfnamefont {X.}~\bibnamefont {Chen}},\ }\href {https://journals.aps.org/prapplied/abstract/10.1103/PhysRevApplied.15.024038} {\bibfield  {journal} {\bibinfo  {journal} {Phys. Rev. Appl.}\ }\textbf {\bibinfo {volume} {15}},\ \bibinfo {pages} {024038} (\bibinfo {year} {2021}{\natexlab{b}})}\BibitemShut {NoStop}%
\bibitem [{\citenamefont {Barkoutsos}\ \emph {et~al.}(2020)\citenamefont {Barkoutsos}, \citenamefont {Nannicini}, \citenamefont {Robert}, \citenamefont {Tavernelli},\ and\ \citenamefont {Woerner}}]{Barkoutsos2020improving}%
  \BibitemOpen
  \bibfield  {author} {\bibinfo {author} {\bibfnamefont {P.~K.}\ \bibnamefont {Barkoutsos}}, \bibinfo {author} {\bibfnamefont {G.}~\bibnamefont {Nannicini}}, \bibinfo {author} {\bibfnamefont {A.}~\bibnamefont {Robert}}, \bibinfo {author} {\bibfnamefont {I.}~\bibnamefont {Tavernelli}},\ and\ \bibinfo {author} {\bibfnamefont {S.}~\bibnamefont {Woerner}},\ }\href {https://doi.org/10.22331/q-2020-04-20-256} {\bibfield  {journal} {\bibinfo  {journal} {{Quantum}}\ }\textbf {\bibinfo {volume} {4}},\ \bibinfo {pages} {256} (\bibinfo {year} {2020})}\BibitemShut {NoStop}%
\bibitem [{\citenamefont {Levinthal}(1968)}]{levinthal1968pathways}%
  \BibitemOpen
  \bibfield  {author} {\bibinfo {author} {\bibfnamefont {C.}~\bibnamefont {Levinthal}},\ }\href {https://doi.org/10.1051/jcp/1968650044} {\bibfield  {journal} {\bibinfo  {journal} {J. Chim. Phys.}\ }\textbf {\bibinfo {volume} {65}},\ \bibinfo {pages} {44} (\bibinfo {year} {1968})}\BibitemShut {NoStop}%
\bibitem [{\citenamefont {Levinthal}(1969)}]{levinthal1969fold}%
  \BibitemOpen
  \bibfield  {author} {\bibinfo {author} {\bibfnamefont {C.}~\bibnamefont {Levinthal}},\ }in\ \href@noop {} {\emph {\bibinfo {booktitle} {{M{\"o}ssbauer Spectroscopy in Biological Systems}}}},\ \bibinfo {editor} {edited by\ \bibinfo {editor} {\bibfnamefont {J.~T.~P.}\ \bibnamefont {DeBrunner}}\ and\ \bibinfo {editor} {\bibfnamefont {E.}~\bibnamefont {M{\"u}nck}}}\ (\bibinfo  {publisher} {University of Illinois Press},\ \bibinfo {address} {Urbana, IL},\ \bibinfo {year} {1969})\ pp.\ \bibinfo {pages} {22--24}\BibitemShut {NoStop}%
\bibitem [{\citenamefont {Jumper}\ \emph {et~al.}(2021)\citenamefont {Jumper}, \citenamefont {Evans}, \citenamefont {Pritzel}, \citenamefont {Green}, \citenamefont {Figurnov}, \citenamefont {Ronneberger}, \citenamefont {Tunyasuvunakool}, \citenamefont {Bates}, \citenamefont {Žídek}, \citenamefont {Potapenko} \emph {et~al.}}]{jumper2021highly}%
  \BibitemOpen
  \bibfield  {author} {\bibinfo {author} {\bibfnamefont {J.}~\bibnamefont {Jumper}}, \bibinfo {author} {\bibfnamefont {R.}~\bibnamefont {Evans}}, \bibinfo {author} {\bibfnamefont {A.}~\bibnamefont {Pritzel}}, \bibinfo {author} {\bibfnamefont {T.}~\bibnamefont {Green}}, \bibinfo {author} {\bibfnamefont {M.}~\bibnamefont {Figurnov}}, \bibinfo {author} {\bibfnamefont {O.}~\bibnamefont {Ronneberger}}, \bibinfo {author} {\bibfnamefont {K.}~\bibnamefont {Tunyasuvunakool}}, \bibinfo {author} {\bibfnamefont {R.}~\bibnamefont {Bates}}, \bibinfo {author} {\bibfnamefont {A.}~\bibnamefont {Žídek}}, \bibinfo {author} {\bibfnamefont {A.}~\bibnamefont {Potapenko}}, \emph {et~al.},\ }\href {https://doi.org/10.1038/s41586-021-03819-2} {\bibfield  {journal} {\bibinfo  {journal} {Nature}\ }\textbf {\bibinfo {volume} {596}},\ \bibinfo {pages} {583} (\bibinfo {year} {2021})}\BibitemShut {NoStop}%
\bibitem [{\citenamefont {Evans}\ \emph {et~al.}(2024)\citenamefont {Evans}, \citenamefont {Mulholland}, \citenamefont {Suliana}, \citenamefont {Pei}, \citenamefont {Qian}, \citenamefont {Senior}, \citenamefont {Yim}, \citenamefont {Jumper}, \citenamefont {Green}, \citenamefont {Ronneberger},\ and\ \citenamefont {Hassabis}}]{evans2024alphafold3}%
  \BibitemOpen
  \bibfield  {author} {\bibinfo {author} {\bibfnamefont {R.}~\bibnamefont {Evans}}, \bibinfo {author} {\bibfnamefont {J.}~\bibnamefont {Mulholland}}, \bibinfo {author} {\bibfnamefont {A.}~\bibnamefont {Suliana}}, \bibinfo {author} {\bibfnamefont {J.}~\bibnamefont {Pei}}, \bibinfo {author} {\bibfnamefont {B.}~\bibnamefont {Qian}}, \bibinfo {author} {\bibfnamefont {A.}~\bibnamefont {Senior}}, \bibinfo {author} {\bibfnamefont {J.}~\bibnamefont {Yim}}, \bibinfo {author} {\bibfnamefont {J.}~\bibnamefont {Jumper}}, \bibinfo {author} {\bibfnamefont {T.}~\bibnamefont {Green}}, \bibinfo {author} {\bibfnamefont {O.}~\bibnamefont {Ronneberger}},\ and\ \bibinfo {author} {\bibfnamefont {D.}~\bibnamefont {Hassabis}},\ }\href {https://doi.org/10.1038/s41586-024-07487-w} {\bibfield  {journal} {\bibinfo  {journal} {Nature}\ }\textbf {\bibinfo {volume} {629}},\ \bibinfo {pages} {319} (\bibinfo {year} {2024})}\BibitemShut {NoStop}%
\bibitem [{\citenamefont {Wang}\ \emph {et~al.}(2022)\citenamefont {Wang}, \citenamefont {Wang}, \citenamefont {Zhang}, \citenamefont {Cheng}, \citenamefont {Yan}, \citenamefont {Shao}, \citenamefont {Wang}, \citenamefont {Wang},\ and\ \citenamefont {Fu}}]{wang2022therapeutic}%
  \BibitemOpen
  \bibfield  {author} {\bibinfo {author} {\bibfnamefont {L.}~\bibnamefont {Wang}}, \bibinfo {author} {\bibfnamefont {N.}~\bibnamefont {Wang}}, \bibinfo {author} {\bibfnamefont {W.}~\bibnamefont {Zhang}}, \bibinfo {author} {\bibfnamefont {X.}~\bibnamefont {Cheng}}, \bibinfo {author} {\bibfnamefont {Z.}~\bibnamefont {Yan}}, \bibinfo {author} {\bibfnamefont {G.}~\bibnamefont {Shao}}, \bibinfo {author} {\bibfnamefont {X.}~\bibnamefont {Wang}}, \bibinfo {author} {\bibfnamefont {R.}~\bibnamefont {Wang}},\ and\ \bibinfo {author} {\bibfnamefont {C.}~\bibnamefont {Fu}},\ }\href {https://doi.org/10.1038/s41392-022-00904-4} {\bibfield  {journal} {\bibinfo  {journal} {Signal Transduction and Targeted Therapy}\ }\textbf {\bibinfo {volume} {7}},\ \bibinfo {pages} {48} (\bibinfo {year} {2022})}\BibitemShut {NoStop}%
\bibitem [{\citenamefont {Dunkelmann}\ \emph {et~al.}(2024)\citenamefont {Dunkelmann}, \citenamefont {Piedrafita}, \citenamefont {Dickson}, \citenamefont {Liu}, \citenamefont {Elliott}, \citenamefont {Fiedler}, \citenamefont {Bellini}, \citenamefont {Zhou}, \citenamefont {Cervettini},\ and\ \citenamefont {Chin}}]{dunkelmann2024adding}%
  \BibitemOpen
  \bibfield  {author} {\bibinfo {author} {\bibfnamefont {D.~L.}\ \bibnamefont {Dunkelmann}}, \bibinfo {author} {\bibfnamefont {C.}~\bibnamefont {Piedrafita}}, \bibinfo {author} {\bibfnamefont {A.}~\bibnamefont {Dickson}}, \bibinfo {author} {\bibfnamefont {K.~C.}\ \bibnamefont {Liu}}, \bibinfo {author} {\bibfnamefont {T.~S.}\ \bibnamefont {Elliott}}, \bibinfo {author} {\bibfnamefont {M.}~\bibnamefont {Fiedler}}, \bibinfo {author} {\bibfnamefont {D.}~\bibnamefont {Bellini}}, \bibinfo {author} {\bibfnamefont {A.}~\bibnamefont {Zhou}}, \bibinfo {author} {\bibfnamefont {D.}~\bibnamefont {Cervettini}},\ and\ \bibinfo {author} {\bibfnamefont {J.~W.}\ \bibnamefont {Chin}},\ }\href {https://doi.org/10.1038/s41586-023-06897-6} {\bibfield  {journal} {\bibinfo  {journal} {Nature}\ }\textbf {\bibinfo {volume} {625}},\ \bibinfo {pages} {603} (\bibinfo {year} {2024})}\BibitemShut {NoStop}%
\bibitem [{\citenamefont {Miura}\ \emph {et~al.}(2023)\citenamefont {Miura}, \citenamefont {Malla}, \citenamefont {Owen}, \citenamefont {Tumber}, \citenamefont {Brewitz}, \citenamefont {McDonough}, \citenamefont {Salah}, \citenamefont {Terasaka}, \citenamefont {Katoh}, \citenamefont {Lukacik} \emph {et~al.}}]{miura2023vitro}%
  \BibitemOpen
  \bibfield  {author} {\bibinfo {author} {\bibfnamefont {T.}~\bibnamefont {Miura}}, \bibinfo {author} {\bibfnamefont {T.~R.}\ \bibnamefont {Malla}}, \bibinfo {author} {\bibfnamefont {C.~D.}\ \bibnamefont {Owen}}, \bibinfo {author} {\bibfnamefont {A.}~\bibnamefont {Tumber}}, \bibinfo {author} {\bibfnamefont {L.}~\bibnamefont {Brewitz}}, \bibinfo {author} {\bibfnamefont {M.~A.}\ \bibnamefont {McDonough}}, \bibinfo {author} {\bibfnamefont {E.}~\bibnamefont {Salah}}, \bibinfo {author} {\bibfnamefont {N.}~\bibnamefont {Terasaka}}, \bibinfo {author} {\bibfnamefont {T.}~\bibnamefont {Katoh}}, \bibinfo {author} {\bibfnamefont {P.}~\bibnamefont {Lukacik}}, \emph {et~al.},\ }\href {https://doi.org/10.1038/s41557-023-01205-1} {\bibfield  {journal} {\bibinfo  {journal} {Nature Chemistry}\ }\textbf {\bibinfo {volume} {15}},\ \bibinfo {pages} {998} (\bibinfo {year} {2023})}\BibitemShut {NoStop}%
\bibitem [{\citenamefont {Miyazawa}\ and\ \citenamefont {Jernigan}(1996)}]{miyazawa1996residue}%
  \BibitemOpen
  \bibfield  {author} {\bibinfo {author} {\bibfnamefont {S.}~\bibnamefont {Miyazawa}}\ and\ \bibinfo {author} {\bibfnamefont {R.~L.}\ \bibnamefont {Jernigan}},\ }\href {https://doi.org/https://doi.org/10.1006/jmbi.1996.0114} {\bibfield  {journal} {\bibinfo  {journal} {Journal of Molecular Biology}\ }\textbf {\bibinfo {volume} {256}},\ \bibinfo {pages} {623} (\bibinfo {year} {1996})}\BibitemShut {NoStop}%
\bibitem [{\citenamefont {Honda}\ \emph {et~al.}(2004)\citenamefont {Honda}, \citenamefont {Yamasaki}, \citenamefont {Sawada},\ and\ \citenamefont {Morii}}]{honda2004residue}%
  \BibitemOpen
  \bibfield  {author} {\bibinfo {author} {\bibfnamefont {S.}~\bibnamefont {Honda}}, \bibinfo {author} {\bibfnamefont {K.}~\bibnamefont {Yamasaki}}, \bibinfo {author} {\bibfnamefont {Y.}~\bibnamefont {Sawada}},\ and\ \bibinfo {author} {\bibfnamefont {H.}~\bibnamefont {Morii}},\ }\href {https://doi.org/https://doi.org/10.1016/j.str.2004.05.022} {\bibfield  {journal} {\bibinfo  {journal} {Structure}\ }\textbf {\bibinfo {volume} {12}},\ \bibinfo {pages} {1507} (\bibinfo {year} {2004})}\BibitemShut {NoStop}%
\bibitem [{\citenamefont {{RCSB Protein Data Bank}}(2003)}]{chignolin}%
  \BibitemOpen
  \bibfield  {author} {\bibinfo {author} {\bibnamefont {{RCSB Protein Data Bank}}},\ }\href {https://www.rcsb.org/structure/1UAO} {\bibinfo {title} {{NMR Structure of designed protein, Chignolin, consisting of only ten amino acids}}} (\bibinfo {year} {2003})\BibitemShut {NoStop}%
\bibitem [{\citenamefont {Bodenmüller}\ and\ \citenamefont {Schaller}(1981)}]{bodenmuller1981conserved}%
  \BibitemOpen
  \bibfield  {author} {\bibinfo {author} {\bibfnamefont {H.}~\bibnamefont {Bodenmüller}}\ and\ \bibinfo {author} {\bibfnamefont {H.~C.}\ \bibnamefont {Schaller}},\ }\href {https://doi.org/10.1038/293579a0} {\bibfield  {journal} {\bibinfo  {journal} {Nature}\ }\textbf {\bibinfo {volume} {293}},\ \bibinfo {pages} {579–580} (\bibinfo {year} {1981})}\BibitemShut {NoStop}%
\bibitem [{\citenamefont {{UniProt Consortium}}(2024{\natexlab{a}})}]{morphneuro}%
  \BibitemOpen
  \bibfield  {author} {\bibinfo {author} {\bibnamefont {{UniProt Consortium}}},\ }\href {https://www.uniprot.org/uniprotkb/P69208/entry} {\bibinfo {title} {{Morphogenetic neuropeptide - Homo sapiens (Human)}}} (\bibinfo {year} {2024}{\natexlab{a}})\BibitemShut {NoStop}%
\bibitem [{\citenamefont {Jackson}\ \emph {et~al.}(2012)\citenamefont {Jackson}, \citenamefont {Wang}, \citenamefont {Gaeta}, \citenamefont {Pomat}, \citenamefont {Siba}, \citenamefont {Rimmer}, \citenamefont {Sewell},\ and\ \citenamefont {Collins}}]{jackson2012divergent}%
  \BibitemOpen
  \bibfield  {author} {\bibinfo {author} {\bibfnamefont {K.~J.~L.}\ \bibnamefont {Jackson}}, \bibinfo {author} {\bibfnamefont {Y.}~\bibnamefont {Wang}}, \bibinfo {author} {\bibfnamefont {B.~A.}\ \bibnamefont {Gaeta}}, \bibinfo {author} {\bibfnamefont {W.}~\bibnamefont {Pomat}}, \bibinfo {author} {\bibfnamefont {P.}~\bibnamefont {Siba}}, \bibinfo {author} {\bibfnamefont {J.}~\bibnamefont {Rimmer}}, \bibinfo {author} {\bibfnamefont {W.~A.}\ \bibnamefont {Sewell}},\ and\ \bibinfo {author} {\bibfnamefont {A.~M.}\ \bibnamefont {Collins}},\ }\href {https://doi.org/10.1007/s00251-011-0559-z} {\bibfield  {journal} {\bibinfo  {journal} {Immunogenetics}\ }\textbf {\bibinfo {volume} {64}},\ \bibinfo {pages} {3–14} (\bibinfo {year} {2012})}\BibitemShut {NoStop}%
\bibitem [{\citenamefont {{UniProt Consortium}}(2024{\natexlab{b}})}]{igjk1}%
  \BibitemOpen
  \bibfield  {author} {\bibinfo {author} {\bibnamefont {{UniProt Consortium}}},\ }\href {https://www.uniprot.org/uniprotkb/A0A0A0MT89/entry} {\bibinfo {title} {{IGKJ1 - Inmunoglobulin kappa joining 1 - Homo sapiens (Human)}}} (\bibinfo {year} {2024}{\natexlab{b}})\BibitemShut {NoStop}%
\bibitem [{\citenamefont {Zielinski}\ \emph {et~al.}(2024)\citenamefont {Zielinski}, \citenamefont {Nublein}, \citenamefont {Kolle}, \citenamefont {Gabor}, \citenamefont {Linnhoff-Popien},\ and\ \citenamefont {Feld}}]{zielinski2024solving}%
  \BibitemOpen
  \bibfield  {author} {\bibinfo {author} {\bibfnamefont {S.}~\bibnamefont {Zielinski}}, \bibinfo {author} {\bibfnamefont {J.}~\bibnamefont {Nublein}}, \bibinfo {author} {\bibfnamefont {M.}~\bibnamefont {Kolle}}, \bibinfo {author} {\bibfnamefont {T.}~\bibnamefont {Gabor}}, \bibinfo {author} {\bibfnamefont {C.}~\bibnamefont {Linnhoff-Popien}},\ and\ \bibinfo {author} {\bibfnamefont {S.}~\bibnamefont {Feld}},\ }in\ \href {https://doi.org/10.1109/QCE60285.2024.00085} {\emph {\bibinfo {booktitle} {2024 IEEE International Conference on Quantum Computing and Engineering (QCE)}}},\ Vol.~\bibinfo {volume} {1}\ (\bibinfo  {publisher} {IEEE Computer Society},\ \bibinfo {address} {Los Alamitos, CA, USA},\ \bibinfo {year} {2024})\ pp.\ \bibinfo {pages} {681--691}\BibitemShut {NoStop}%
\bibitem [{\citenamefont {Philathong}\ \emph {et~al.}(2021)\citenamefont {Philathong}, \citenamefont {Akshay}, \citenamefont {Samburskaya},\ and\ \citenamefont {Biamonte}}]{Philathong_2021}%
  \BibitemOpen
  \bibfield  {author} {\bibinfo {author} {\bibfnamefont {H.}~\bibnamefont {Philathong}}, \bibinfo {author} {\bibfnamefont {V.}~\bibnamefont {Akshay}}, \bibinfo {author} {\bibfnamefont {K.}~\bibnamefont {Samburskaya}},\ and\ \bibinfo {author} {\bibfnamefont {J.}~\bibnamefont {Biamonte}},\ }\href {https://doi.org/10.1088/2632-072X/abdadc} {\bibfield  {journal} {\bibinfo  {journal} {Journal of Physics: Complexity}\ }\textbf {\bibinfo {volume} {2}},\ \bibinfo {pages} {011002} (\bibinfo {year} {2021})}\BibitemShut {NoStop}%
\bibitem [{\citenamefont {Gent}\ and\ \citenamefont {Walsh}(1994)}]{gent1994sat}%
  \BibitemOpen
  \bibfield  {author} {\bibinfo {author} {\bibfnamefont {I.~P.}\ \bibnamefont {Gent}}\ and\ \bibinfo {author} {\bibfnamefont {T.}~\bibnamefont {Walsh}},\ }in\ \href {https://dl.acm.org/doi/10.5555/3070217.3070238} {\emph {\bibinfo {booktitle} {Proceedings of the 11th European Conference on Artificial Intelligence}}},\ \bibinfo {series and number} {ECAI'94}\ (\bibinfo  {publisher} {John Wiley \& Sons, Inc.},\ \bibinfo {address} {USA},\ \bibinfo {year} {1994})\ p.\ \bibinfo {pages} {105–109}\BibitemShut {NoStop}%
\bibitem [{\citenamefont {Kirkpatrick}\ and\ \citenamefont {Selman}(1994)}]{kirkpatrick1994critical}%
  \BibitemOpen
  \bibfield  {author} {\bibinfo {author} {\bibfnamefont {S.}~\bibnamefont {Kirkpatrick}}\ and\ \bibinfo {author} {\bibfnamefont {B.}~\bibnamefont {Selman}},\ }\href {https://doi.org/10.1126/science.264.5163.1297} {\bibfield  {journal} {\bibinfo  {journal} {Science}\ }\textbf {\bibinfo {volume} {264}},\ \bibinfo {pages} {1297} (\bibinfo {year} {1994})}\BibitemShut {NoStop}%
\bibitem [{\citenamefont {Ignatiev}\ \emph {et~al.}(2024)\citenamefont {Ignatiev}, \citenamefont {Tan},\ and\ \citenamefont {Karamanos}}]{ignatiev2024towards}%
  \BibitemOpen
  \bibfield  {author} {\bibinfo {author} {\bibfnamefont {A.}~\bibnamefont {Ignatiev}}, \bibinfo {author} {\bibfnamefont {Z.~L.}\ \bibnamefont {Tan}},\ and\ \bibinfo {author} {\bibfnamefont {C.}~\bibnamefont {Karamanos}},\ }in\ \href {https://doi.org/10.4230/LIPIcs.SAT.2024.16} {\emph {\bibinfo {booktitle} {27th International Conference on Theory and Applications of Satisfiability Testing (SAT 2024)}}},\ \bibinfo {series} {Leibniz International Proceedings in Informatics (LIPIcs)}, Vol.\ \bibinfo {volume} {305},\ \bibinfo {editor} {edited by\ \bibinfo {editor} {\bibfnamefont {S.}~\bibnamefont {Chakraborty}}\ and\ \bibinfo {editor} {\bibfnamefont {J.-H.~R.}\ \bibnamefont {Jiang}}}\ (\bibinfo  {publisher} {Schloss Dagstuhl -- Leibniz-Zentrum f{\"u}r Informatik},\ \bibinfo {address} {Dagstuhl, Germany},\ \bibinfo {year} {2024})\ pp.\ \bibinfo {pages} {16:1--16:11}\BibitemShut {NoStop}%
\bibitem [{\citenamefont {PySAT}(2021)}]{pysat}%
  \BibitemOpen
  \bibfield  {author} {\bibinfo {author} {\bibnamefont {PySAT}},\ }\href@noop {} {\bibinfo {title} {{A toolkit for SAT-based prototyping in Python}}},\ \bibinfo {howpublished} {\url{https://github.com/pysathq/pysat}} (\bibinfo {year} {2021})\BibitemShut {NoStop}%
\bibitem [{\citenamefont {Metropolis}\ \emph {et~al.}(1953)\citenamefont {Metropolis}, \citenamefont {Rosenbluth}, \citenamefont {Rosenbluth}, \citenamefont {Teller},\ and\ \citenamefont {Teller}}]{metropolis1953equation}%
  \BibitemOpen
  \bibfield  {author} {\bibinfo {author} {\bibfnamefont {N.}~\bibnamefont {Metropolis}}, \bibinfo {author} {\bibfnamefont {A.~W.}\ \bibnamefont {Rosenbluth}}, \bibinfo {author} {\bibfnamefont {M.~N.}\ \bibnamefont {Rosenbluth}}, \bibinfo {author} {\bibfnamefont {A.~H.}\ \bibnamefont {Teller}},\ and\ \bibinfo {author} {\bibfnamefont {E.}~\bibnamefont {Teller}},\ }\href {https://doi.org/10.1063/1.1699114} {\bibfield  {journal} {\bibinfo  {journal} {The Journal of Chemical Physics}\ }\textbf {\bibinfo {volume} {21}},\ \bibinfo {pages} {1087} (\bibinfo {year} {1953})}\BibitemShut {NoStop}%
\bibitem [{\citenamefont {Hastings}(1970)}]{hastings1970monte}%
  \BibitemOpen
  \bibfield  {author} {\bibinfo {author} {\bibfnamefont {W.~K.}\ \bibnamefont {Hastings}},\ }\href {https://doi.org/10.1093/biomet/57.1.97} {\bibfield  {journal} {\bibinfo  {journal} {Biometrika}\ }\textbf {\bibinfo {volume} {57}},\ \bibinfo {pages} {97} (\bibinfo {year} {1970})}\BibitemShut {NoStop}%
\bibitem [{\citenamefont {Chen}\ \emph {et~al.}(2024)\citenamefont {Chen}, \citenamefont {Nielsen}, \citenamefont {Ebert}, \citenamefont {Inlek}, \citenamefont {Wright}, \citenamefont {Chaplin}, \citenamefont {Maksymov}, \citenamefont {Páez}, \citenamefont {Poudel}, \citenamefont {Maunz} \emph {et~al.}}]{Chen2024-co}%
  \BibitemOpen
  \bibfield  {author} {\bibinfo {author} {\bibfnamefont {J.-S.}\ \bibnamefont {Chen}}, \bibinfo {author} {\bibfnamefont {E.}~\bibnamefont {Nielsen}}, \bibinfo {author} {\bibfnamefont {M.}~\bibnamefont {Ebert}}, \bibinfo {author} {\bibfnamefont {V.}~\bibnamefont {Inlek}}, \bibinfo {author} {\bibfnamefont {K.}~\bibnamefont {Wright}}, \bibinfo {author} {\bibfnamefont {V.}~\bibnamefont {Chaplin}}, \bibinfo {author} {\bibfnamefont {A.}~\bibnamefont {Maksymov}}, \bibinfo {author} {\bibfnamefont {E.}~\bibnamefont {Páez}}, \bibinfo {author} {\bibfnamefont {A.}~\bibnamefont {Poudel}}, \bibinfo {author} {\bibfnamefont {P.}~\bibnamefont {Maunz}}, \emph {et~al.},\ }\href {https://doi.org/10.22331/q-2024-11-07-1516} {\bibfield  {journal} {\bibinfo  {journal} {Quantum}\ }\textbf {\bibinfo {volume} {8}},\ \bibinfo {pages} {1516} (\bibinfo {year} {2024})}\BibitemShut {NoStop}%
\bibitem [{\citenamefont {{Kim}}\ \emph {et~al.}(2008)\citenamefont {{Kim}}, \citenamefont {{McLeod}}, \citenamefont {{Saffman}},\ and\ \citenamefont {{Wagner}}}]{Kim:2008ApOpt}%
  \BibitemOpen
  \bibfield  {author} {\bibinfo {author} {\bibfnamefont {S.}~\bibnamefont {{Kim}}}, \bibinfo {author} {\bibfnamefont {R.~R.}\ \bibnamefont {{McLeod}}}, \bibinfo {author} {\bibfnamefont {M.}~\bibnamefont {{Saffman}}},\ and\ \bibinfo {author} {\bibfnamefont {K.~H.}\ \bibnamefont {{Wagner}}},\ }\href {https://doi.org/10.1364/AO.47.001816} {\bibfield  {journal} {\bibinfo  {journal} {Appl. Opt.}\ }\textbf {\bibinfo {volume} {47}},\ \bibinfo {pages} {1816} (\bibinfo {year} {2008})}\BibitemShut {NoStop}%
\bibitem [{\citenamefont {{Pogorelov}}\ \emph {et~al.}(2021)\citenamefont {{Pogorelov}}, \citenamefont {{Feldker}}, \citenamefont {{Marciniak}}, \citenamefont {{Postler}}, \citenamefont {{Jacob}}, \citenamefont {{Krieglsteiner}}, \citenamefont {{Podlesnic}}, \citenamefont {{Meth}}, \citenamefont {{Negnevitsky}}, \citenamefont {{Stadler}} \emph {et~al.}}]{Pogorelov:2021PRXQ}%
  \BibitemOpen
  \bibfield  {author} {\bibinfo {author} {\bibfnamefont {I.}~\bibnamefont {{Pogorelov}}}, \bibinfo {author} {\bibfnamefont {T.}~\bibnamefont {{Feldker}}}, \bibinfo {author} {\bibfnamefont {C.~D.}\ \bibnamefont {{Marciniak}}}, \bibinfo {author} {\bibfnamefont {L.}~\bibnamefont {{Postler}}}, \bibinfo {author} {\bibfnamefont {G.}~\bibnamefont {{Jacob}}}, \bibinfo {author} {\bibfnamefont {O.}~\bibnamefont {{Krieglsteiner}}}, \bibinfo {author} {\bibfnamefont {V.}~\bibnamefont {{Podlesnic}}}, \bibinfo {author} {\bibfnamefont {M.}~\bibnamefont {{Meth}}}, \bibinfo {author} {\bibfnamefont {V.}~\bibnamefont {{Negnevitsky}}}, \bibinfo {author} {\bibfnamefont {M.}~\bibnamefont {{Stadler}}}, \emph {et~al.},\ }\href {https://doi.org/10.1103/PRXQuantum.2.020343} {\bibfield  {journal} {\bibinfo  {journal} {PRX Quantum}\ }\textbf {\bibinfo {volume} {2}},\ \bibinfo {eid} {020343} (\bibinfo {year} {2021})}\BibitemShut {NoStop}%
\end{thebibliography}%

\end{document}